\documentclass[11pt]{article}
\usepackage[T1]{fontenc}
\usepackage[utf8]{inputenc}
\usepackage{lmodern}
\usepackage{upgreek}
\usepackage{jheppub}
\usepackage{subcaption}
\usepackage{ragged2e}
 
\usepackage{amsmath,amssymb,amsfonts,dsfont,mathrsfs,amsthm,mathtools,bbm}
\definecolor{rossocorsa}{rgb}{0.83, 0.0, 0.0}
\definecolor{bleudefrance}{rgb}{0.19, 0.55, 0.91}

\usepackage{hyperref}
\hypersetup{
    colorlinks,
    citecolor=bleudefrance,
    filecolor=bleudefrance,
    linkcolor=rossocorsa,
    urlcolor=bleudefrance
}

\newcommand{\arxiv}[1]{{\tt
\href{http://www.arXiv.org/abs/#1}{#1}}}
\newcommand{\doi}[1]{{\tt DOI:\href{https://doi.org/#1}{#1}}}

\begin{document}

\title{\Huge Photon spheres and bulk probes in $\text{AdS}_3/\text{CFT}_2$:\,the quantum BTZ black hole}

\author[a]{Oscar Lasso Andino,}
\author[b]{Axel Le\'{o}n-Arteaga,}
\author[b]{Guillermo Ram\'{i}rez-Ulloa ,}

\affiliation[a]{Escuela de Ciencias Físicas y Matemáticas, Universidad de Las Américas,\\ Redondel del ciclista, Antigua vía a Nayón, C.P. 170504, Quito, Ecuador
\vspace{0.1cm}
}
\affiliation[b]{Departamento de F\'{i}sica, Universidad Nacional de Trujillo, Avenida Juan Pablo II S/N; Ciudad Universitaria, Trujillo, La Libertad, Per\'{u}}

\emailAdd{oscar.lasso@udla.edu.ec}
\emailAdd{axleon@unitru.edu.pe}
\emailAdd{gramirezu@unitru.edu.pe}

\abstract{
 The entanglement entropy in $d+1$ dimensional conformal field theories can be calculated using the area of $d$ dimensional minimal surfaces in $AdS_{d+2}$.
Therefore, the existence of surfaces anchored in the boundary of an asymptotically anti-de Sitter (AdS) spacetime is crucial for the calculation of entanglement entropy. In particular, in $d=3$ the extremal surfaces are geodesics with two ends in the boundary. In the Schwarzschild-AdS black hole the space-like geodesics can  connect timelike-separated points by winding around the horizon multiple times. This result can be extended to other asymptotically AdS spacetimes. For geodesics joining time-like separated points, if there is a photon ring then the timelike entanglement entropy in the $\text{AdS}_3/\text{CFT}_2$ will not have an imaginary part.  We present an exhaustive analysis about the existence of geodesics anchored in the boundary of the three dimensional quantum BTZ (quBTZ) black hole and its charged counterpart. We found conditions for the existence of geodesics with two ends in the boundary in all branches of the quBTZ and determine the type of distance between the points in the boundary.  We use a criteria for the existence of light rings to shed some light over the conjecture for spacetimes that are spherically symmetric and have a photon sphere: there are always points with time-like separation that can be connected by space-like or null geodesics.
}

\maketitle


\setlength\parindent{0pt}

\pagestyle{plain}


\newpage


\section{Introduction}

The existence of a duality between Anti-de Sitter  spacetime and a conformal field theory that resides in the boundary of it has been one of the most powerful, and hence, developed ideas in the high energy physics community in the last years. The AdS/CFT duality describes a correspondence between a $(d+2)$ - dimensional gravitational theory defined in an AdS spacetime and a $(d+1)$-dimensional conformal, non-gravitational, field theory located in the boundary of that AdS spacetime \cite{Maldacena:1997re,Witten:1998qj}. The AdS/CFT duality provides a way to study strongly coupled systems in the boundary or the properties of black holes in the bulk. On this direction, the AdS/CFT duality offers a promising way of studying the properties of spacetimes at very small scales, namely a theory of quantum gravity.\\
A correlation function in a CFT defined on the boundary  can be calculated using geodesics defined in the bulk. One preferred method is the geodesic approximation \cite{Louko:2000tp,Aalsma:2022eru}. Thus, a correlation function, which is a complex quantum field theory calculation, can be elucidated  by setting a geometry problem, namely the length of a shortest path  connecting two point sin the boundary. Therefore, a study of the existence of geodesics anchored in the boundary is needed before calculating correlation functions in the CFT. Moreover, in $d=3$ the study of  geodesics anchored in the boundary is of the paramount importance, it is because in the Ryu-Takayanagi prescription \cite{Ryu:2006bv} for the calculation of entanglement entropy the extremal surfaces are geodesics in the bulk. Therefore it could happen that the Ryu-Takayanagi formula cannot be used for calculation of entanglement entropy.\\
On the other hand, the idea that time can be an emergent property comming from a fundamental quantum information property motivates the concept of timelike entanglement entropy \cite{
Doi:2022iyj,Doi:2023zaf,Nakata:2020luh,Li:2022tsv,Anegawa:2024kdj,Guo:2025pru,Roychowdhury:2025ebs, Heller:2024whi}.  In $\text{AdS}_3$ the existence of geodesics with both ends anchored in the boundary is pivotal for the calculation of timelike entanglement entropy. This generalization of the entanglement entropy can only be calculated by knowing the length of geodesics from the boundary to the turning point. Even when the existence of a geodesic with both ends anchored in the boundary  is secured, the analysis of the distance between both time-like points in the boundary is not trivial. The analysis  of these type of geodesics is important for entanglement entropy calculations, the solely existence of geodesics anchored in the boundary in points with time-like separation is an important problem by itself \cite{Engelhardt:2013tra}.
In \cite{He:2024emd} it has been shown that in Schwarzschild-AdS (SAdS) spacetimes there are bulk geodesics that connect arbitrary two points on the boundary with timelike separation, the AdS and BTZ cases where also sudied. In this context, one of the objectives of this article is to study the existence of geodesics anchored in the boundary connecting time-like points. We primarily focus on a generalization of the well known BTZ black hole metric \cite{Banados:1992wn}, namely the quantum BTZ (qubTZ) black hole metric  and its charged counterpart \cite{Emparan:2020znc,Emparan:2002px, Emparan:1999fd,Feng:2024uia}, see also \cite{Chen:2023tpi,Frassino:2024bjg} for some properties of quBTZ black holes.\\
We define three types of geodesics anchored in the boundary \cite{Hubeny:2006yu}.
Geodesics that only have one point in the boundary are called of the $Type\,\,\mathcal{B}$, geodesics lying entirely in the boundary are of the $Type\,\,\mathcal{A}$. Finally, geodesics with both endpoints in the boundary are of the $Type\,\,\mathcal{C}$. If the two endpoints of a  geodesic lie on different boundaries, we also classify it  as a $Type\,\,\mathcal{B}$ geodesic. Our interest lies in the $Type\,\,\mathcal{C}$ geodesics. This type of geodesics are the ones that correspond to extremal surfaces in a $2+1$ -dimensional asymptotically $\text{AdS}$ spacetime.  However, being of $Type\,\,\mathcal{C}$ is not enough, as we are going to show, the causal structure of endpoints is the uttermost importance.\\ 
We also want to shed some light on the conjecture that states that if a spacetime has a photon sphere/ring then it is always possible to find a geodesic connecting a pair of time-like points on the boundary \cite{He:2024emd, Hashimoto:2019jmw}. This conjecture is related to a deeper question regarding the relationship between light rings and the CFT defined in the boundary of an asymptotically AdS spacetime \cite{Hashimoto:2019jmw, Hashimoto:2018okj, Liu:2022cev}. In order to attack the problem we relay on a purely geometric method that uses the Gaussian and geodesic curvatures  of surfaces built by projecting the spacetime metric over constant energy surfaces. These new geometric methods are used for studying photon spheres and massive particle surfaces \cite{Qiao:2022jlu,Qiao:2022hfv,Bermudez-Cardenas:2024bfi,Bermudez-Cardenas:2025duw, Arganaraz:2021fwu}. The method uses a Riemanian metric obtained by dimensional reduction of a spacetime metric  and uses intrinsic curvatures of the Riemannian metric for classifying circular trajectories and study its stability. The procedure has been applied to different spacetimes and generalizations towards stationary and asymptotically AdS spacetimes have been carried out \cite{Bermudez-Cardenas:2025duw, Bermudez-Cardenas:2025hrp, Arganaraz:2021fwu}.\\ 
Using the curvatures of the Jacobi metric we proceed to study the existence of geodesics of $Type\,\,\mathcal{C}$. The conditions for the existence of this type of geodesics, coming from the effective potential of the radial geodesic equations, are translated to conditions over the Gaussian and geodesic curvatures of the Jacobi metric.\\
In section \ref{sec:2} we present a review of the formulae regarding the geodesic probes using the effective potential. We summarize the results known for pure AdS, Schwarzschild-AdS and set up the notation. In section \ref{sec:3} we study the geodesic probes of the quBTZ black hole. We analyze all different geometries. Due to the fact that there are not time-like geodesics connecting two points in the boundary we study all possibilites for null and space-like geodesics. We also analyze the causality between the points in the boundary that are connected by a geodesic in the bulk.  In section \ref{sec:4} we analyze geodesic probes for the charged quantum BTZ black hole. In section \ref{sec:5} we provide a simple a argument for the existence of turning points when there is a photon sphere in the bulk, giving some light to a previously conjectured relationship. In section \ref{sec:6} we present the conclusions section.

\section{Geodesics anchored in the boundary}\label{sec:2}
The existence of geodesics that connect points in the boundary is an important issue that needs to be clarified before applying any geometric approach for calculating the entanglement entropy. A particular example is the calculation of correlation functions in the boundary, using geodesic approximation they can be carried out only using geometry. It implies that the problem of existence of geodesics connecting two points (local operators) in the boundary needs to be solved,  before using the Ryu-Takayangi prescription for example, for calculating entanglement entropy. \\
The three types of geodesics that are going to be considered are: the ones that lay completely in the bulk (Type $\,\mathcal{A}$). The ones that have one end in the boundary $(Type \,\mathcal{B}$) and geodesics that have both ends in the boundary  $(Type \,\mathcal{C}$) \footnote{In spacetime geometries that have several boundaries it could happen that a geodesics have one end in one boundary and the other end in the other boundary, these type of geodesics are classified as of the $(Type \,\mathcal{B}$). } The geodesics of $(Type \,\mathcal{C}$) correspond  to extremal surfaces in $2+1$-dimensional spacetime. Using the potential appearing in the geodesic equation, the problem of determining the existence of geodesics connecting points in the boundary can be treated as an initial value problem. In other words, the boundary value problem is transformed to a problem  of motion of free particles whose starting point is located at the boundary. The idea is to find a condition for the existence of a potential barrier that reflects the particle. Moreover, the method  provides a way to study the causal structure of the endpoints lying on the boundary. Let us show how the method works.  Consider a metric of the type 
\begin{equation}\label{m:1}
ds^2=-f(r)dt^2+f^{-1}(r)dr^2+r^2 d\Omega^2
\end{equation}
where $f(r)\rightarrow r^2+1$ as $r \rightarrow \infty$, the metric \eqref{m:1} is asymptotically AdS and spherically symmetric. Then by fixing angular coordinates we can confine a geodesic to the equator of a sphere. The tangent vector to that geodesic can be written \cite{Hubeny:2012ry}
\begin{equation}\label{pa:1}
p^{a}=\dot{t}\partial^a_t+\dot{r}\partial^a_r+\dot{\phi}\partial^a_{\phi},
\end{equation}
where $\dot{t}=\frac{dt}{d\lambda}$, with $\lambda$ being the afine parameter of the geodesics. The radial geodesic equation can be written as 
\begin{equation}\label{geor}
\dot{r}^2+V_{\rm eff}(r)=0,\,\,\,\,\,V_{\rm eff}=f(r)\left(-\mathcal{K} -\frac{E^2}{f(r)}+\frac{L^2}{r^2}\right),
\end{equation}
where $\mathcal{K}=p^{a}p_{a}=0,\pm 1$ and the usual constants of motion, the energy $E$ and the angular momentum $L$ are given by
\begin{equation}\label{EL:1}
E=-p_{a}\partial_{t}^{a}=f(r)\dot{t},\,\,\,L=p_{a}\partial_{\phi}^{a}=\dot{r}^2\dot{\phi}.
\end{equation}
Geodesic motion requires $\dot{r}^2\geq 0$, therefore from \eqref{geor} we conclude that the only condition needed for a geodesic touching the boundary is $\lim_{r\rightarrow \infty}V_{\rm eff}(r)<0$. Moreover, the point inside the bulk which constitutes a turning point $r_t$ is the largest root of $V_{\rm eff}(r)$.\\
The effective potential defined in \eqref{geor} can be rewritten as
\begin{equation}\label{rad:1}
V_{\rm eff}=-\frac{\mathcal{K}f(r)}{L^2}+\frac{f(r)}{r^2}-\frac{1}{\sigma^2},
\end{equation}
where $\sigma=L/E$  is called the impact parameter. Note that in order to determine the effective potential we have to make the scaling $\lambda=\frac{\lambda}{L}$ in geodesic equation \eqref{geor}. In the case of null geodesics, the parameter  $\sigma$ can be used to characterize the geodesic completely, however in the timelike and spacelike cases we must use $\sigma$ and $L$ as the geodesic parameters. Without losing generality we are going to consider only non-negative values of $E$ and $L$.\\
The negativity of the potential \eqref{rad:1} at infinity together with the existence of points $r_{t}$ such that $V_{\rm eff}(r_t)=0$ leads to the existence of geodesic of $Type \,\mathcal{C}$. It is clear that a geodesic that has one end at the boundary has two options: either it goes inside the bulk without coming back or there is a point where the geodesic returns to the boundary, and therefore becomes a geodesic of $Type\,\mathcal{C}$. However, the points in the boundary that are joined by a geodesic of $Type\,\mathcal{C}$ are causally connected, therefore it is important to determine the type of causality. Due to the fact that we are working in three dimensions, a calculation in the boundary will involve only two coordinates, namely $t$ and $\phi$. Thus, the sign of the interval defined in the boundary will tell us if the separation between both points is time-like, space-like or null. We need to calculate the temporal distance $\Delta t$ and the angular distance $\Delta \phi$ between geodesic endpoints. From equation \eqref{rad:1} we obtain

\begin{eqnarray}
\Delta t&=&\int_{r_{t}}^{\infty}\frac{r}{f(r)\sqrt{\frac{\mathcal{K} r^2 f
(r)}{E^2}+r^2-\sigma^2f(r)}}dr\label{dt}\\
\Delta \phi&=&\int_{r_t}^{\infty}\frac{\sigma}{r^2\sqrt{\frac{\mathcal{K} r^2 
f(r)}{E^2}+r^2-\sigma^2f(r)}}dr\label{dp},
\end{eqnarray}

We are interested in the time-like causal connection, this condition in a two dimensional Lorentzian spacetime can be written as \cite{Gao:2000ga} $\Delta t\geq \Delta \phi$ and it can be translated to the following inequality
\begin{equation}
\left\vert\frac{\Delta t}{\Delta \phi}\right\vert\geq 1
\end{equation} 
where $\Delta t=-2t(r_t)$ and $\Delta \phi=-2\phi(r_t)$. In order to account for the total geodesic distance, we have doubled the distance from the turning point to the boundary. \\
Let us show how the previous procedure works for AdS spacetime. The pure AdS metric in three dimensions can be written as
\begin{equation}
ds^2=-f(r)dt^2+\frac{1}{f(r)}dr^2+r^2d\mathcal{X}^2_{\kappa,2}
\end{equation}
where 
\begin{equation}\label{ads}
f(r)=\kappa+\frac{r^2}{\ell^2},
\end{equation}
with $\kappa=0,\pm 1$. The AdS radius $\ell$ can be set to one. The radial coordinate is $r$, the boundary is located at $r=+\infty$ and $d\mathcal{X}^2_{\kappa,2}$  is given by

\begin{gather*}
d\mathcal{X}^2_{\kappa,2}=
\begin{cases}
 d\Omega_2^2=d\theta^2+\sin^2(\theta)d\Omega_{1}^2,\,\,\,\,for\,\,\kappa=+1, \\
dX_2^2=dx_1^2+dx_2^2,\,\,\,\,\,\,\,\,\,\,\,\,\,\,\,\,\,\,\,\,for\,\,\kappa=0, \\
d\Theta^2_2=d\theta^2+\sinh^2 \theta d\Omega_1^2,\,\,\,for\,\,\kappa=-1.
\end{cases}
\end{gather*}

The effective potential $V_{\rm eff}$ in equation \eqref{rad:1} for the AdS metric \eqref{ads} becomes

\begin{equation}\label{pot:ads}
V_{\rm eff}=\frac{\kappa}{r^2}+1-\frac{1}{\sigma^2}-\mathcal{K}\frac{\kappa +r^2}{L^2}.
\end{equation}
From the previous equation it is direct to see that for time-like geodesics $(\mathcal{K}=-1)$  we have $\lim_{r\rightarrow \infty}V_{\rm eff}=+\infty$, therefore according to the criteria above the radial geodesic equation \eqref{rad:1} has not solutions with at least one end anchored in the boundary. The only possibilities are null and space-like geodesics.\\
For null geodesics $\mathcal{K}=0$, then from \eqref{pot:ads} we can conclude that at infinity the potential is going to be negative if and only if $1-\frac{1}{\sigma^2}>0$, therefore  independently  of the value of $\kappa$, there is going to be a fixed geodesic on the boundary when $\sigma \in (0,1]$. However, when $\kappa=1$ (spherical case) the potential $V_{\rm eff}\rightarrow+\infty$ when $r\rightarrow 0$. Then, if the geodesic is able to return there must be a point where $\dot{r}=0$ which implies that $V_{\rm eff}=0$ at $r_t=\pm \sigma/\sqrt{1-\sigma^2}$, we conclude that there is always a null geodesic of the $Type \,\, \mathcal{C}$.\\
In order to study  causality between points connected by these geodesics we $\Delta t$ and $\Delta \phi$ defined in \eqref{dt} and  \eqref{dp}.  It can be shown that  $\big|\frac{\Delta t}{\Delta \phi}\big|=1$, then  two points joined by null geodesics of the $Type \,\, \mathcal{C}$ have light-like separation. A similar analysis can be done for the remaining geometries $\kappa=0$ and $\kappa=-1$. Indeed, when $\kappa =0$, null geodesics with $\sigma<1$ never turn back, extending towards $r=0$. When $\sigma=1$ there are only geodesics of  $Type \,\, \mathcal{A}$. Finally, when $\kappa=-1$, null geodesics with an endpoint in the boundary at $r=\infty$ pass through $r=0$ towards $r=-\infty$, therefore we will have only null geodesics of $Type \,\, \mathcal{B}$.\\
A similar analysis can be carried out for spacelike geodesics $(\mathcal{K}=1)$. Thus, for spherical symmetry $\kappa =1$ the behavior of geodesics is very similar to the null case and spherical symmetry, where all geodesics departing from the boundary return to the boundary if $\sigma\in (0,\infty)$. Following the usual procedure it can be shown that $\big|\frac{\Delta t}{\Delta \phi}\big|\leq 1$, then spacelike geodesics of  $Type,\,\mathcal{C}$ can only connect spacelike points. When, $\kappa=0$ then $\big|\frac{\Delta t}{\Delta \phi}\big|\leq 1$. Finally, for hyperbolic geometry $(\kappa=-1)$ there are spacelike geodesics of $Type\,\,\mathcal{C}$ for certain fixed value of $\sigma$. When $L<1$ the turning point of the spacelike geodesic lies inside the horizon and therefore, the geodesic will not return to the boundary. A detailed discussion for pure AdS  and other three dimensional spacetimes can be found in  \cite{He:2024emd}. In the next section we study the existence of geodesics of $Type,\,\mathcal{C}$ of the quBTZ black holes.\\

\section{The quantum BTZ black hole}\label{sec:3}

The quantum BTZ black hole (quBTZ) is  built by solving a semi-classical version of Einstein equations. The quBTZ black hole is localized in a brane in $AdS_4$ and it was found  by studying a classical bulk dual to a black hole localized on a brane. These black holes can be interpreted as the duals of certain quantum corrected black holes in $d-$dimensions. A detailed study of its construction, its properties and several applications can be found in \cite{Emparan:2020znc,Emparan:2002px,Emparan:1999fd,Cui:2025qdy,Xu:2024iji,Climent:2024nuj,Wu:2024txe,HosseiniMansoori:2024bfi,Johnson:2023dtf}. In the static, non-rotating case  the metric of a quBTZ is given by 
\begin{equation}\label{qBTZ:1}
ds^2=-f(r)dt^2+\frac{dr^2}{f(r)}+r^2d\phi^2, 
\end{equation}
where 
\begin{equation}\label{f:1}
f(r)=\left(\frac{r^2}{\ell^2_3}-8\mathcal{G}_3 M-\frac{\ell F(M)}{r}\right)
\end{equation}
and $\mathcal{G}_3$ is the renormalized Newton constant\footnote{The renormalized Newton constant is written as
\begin{equation*}
\mathcal{G}_3=\frac{G_3}{\sqrt{1+(\ell/\ell_3)^2}}
\end{equation*}. However, we use the expression 
\begin{equation*}
\mathcal{G}_3=\frac{\ell_4}{\ell}G_3
\end{equation*}  
which is equivalent up to $\mathcal{O}(\ell/L_3)^4$.}, $\ell_{3}$ is the curvature radius of the brane, with $M$ being the mass, $\ell$ the infra-red cut-off length and with $F(M)$ given by \footnote{The variable $x_1$ represents the smallest root of the function $G(x)=1-\kappa x^2-\mu x^3$. The function $G(x)$ appears in the $AdS_4$ C-metric.}
 
\begin{equation}\label{F:1}
F(M)=8\frac{1-\kappa x_1^2}{(3-\kappa x_1^2)^3}.
\end{equation}
When $\ell =0$ the classical BTZ solution is recovered. The $3-$dimensional mass is given by
\begin{equation}\label{mqbtz}
M=-\frac{1}{2G_3}\frac{\ell}{\ell_4}\frac{\kappa x_1^2}{(3-\kappa x_1^2)^2},
\end{equation}
which is bounded from above and below:
\begin{equation}
-\frac{1}{8\mathcal{G}_3}\leq M\leq \frac{1}{24 \mathcal{G}_3}.
\end{equation}

The values $\kappa x_{1}^2=\pm 1,0$ lead to different solutions (brane slicings)\footnote{When $\kappa=-1$ the classical BTZ black hole is recovered}. The branch \textbf{1a} is defined by $\kappa=+1$ and $0<x_1<1$ leading to negative mass. When $\kappa=-1$ we have positive mass and two branches. Branch \textbf{1b} is defined by $0<x_1<\sqrt{3}$ and branch \textbf{2} when $\sqrt{3}<x_1<\infty$, and both branches meet at $\sqrt{3}$. However, all physical quantities depend on $\kappa x_1^2$, not on $\kappa$  and $x_1^2$ separately, therefore it is customary to define two branches
\begin{eqnarray}
Branch\,\, 1&:&\,\,\,-1<-\kappa x_1^2<3,\\
Branch\,\, 2&:&\,\,\,\,\,\,\,\,3<-\kappa x_1^2<\infty
\end{eqnarray}

 The interesting fact is that classically, all solutions exist in different branches completely disconnected, however when quantum effects are considered all branches are unified. Note that the quantum correction is not a Planckian effect. 
The metric \eqref{qBTZ:1} is called quantum because it is a solution to the semiclassical Einstein equations (on the brane) at all orders in quantum back reaction, whose strength is controlled by $\ell$. Although the metric \eqref{qBTZ:1} can be written in other coordinates \footnote{The coordinates on which the metric \eqref{qBTZ:1} is written are not very useful when studying  event horizons. Other coordinates can be introduced using the quotient $\ell/\ell_3$ and $\mu=(1-\kappa x_1^2)/x_1^3$ lie in the interval on which the roots of $H(r)$ are located, in the following way:
\begin{equation}\label{cordn:1}
z=\frac{\ell_3}{r_{+}\,x_1},\,\,\,\,\, \nu=\frac{\ell}{\ell_3},
\end{equation}
then
\begin{eqnarray}
x_{1}^{2}&=&-\frac{1}{\kappa}\frac{1-\nu z^3}{z^2 (1+\nu z)},\\
r_{+}^2&=&-\ell^2_3 \kappa\frac{1+\nu z}{1-\nu z^3},\\
\mu x_1&=&-\kappa \frac{1+z^2}{1-\nu z^3}.
\end{eqnarray}
where $r_{+}$ is the location of the horizon.} we are going stick to the ones that depend explicitly on $\kappa x_1$.
Replacing \eqref{F:1} and \eqref{mqbtz} in \eqref{f:1} we can write
\begin{equation}\label{f:2}
f(r)=\left(r^2+\frac{4\kappa x_1^2}{(3-\kappa x_1^2)^2}-\frac{8(1-\kappa x_1^2)}{r(3-\kappa x_1^2)^3}\right).
\end{equation}
In the following we are going to use the metric function \eqref{f:2} in all calculations in three cases, namely $kx_1^2=0,\pm 1$. The effective potential \eqref{rad:1} for the metric function \eqref{f:2} is going to be written
\begin{equation}\label{pot:fin}
V_{\rm eff}=-\left(r^2+\frac{4\kappa x_1^2}{(3-\kappa x_1^2)^2}-\frac{8(1-\kappa x_1^2)}{r(3-\kappa x_1^2)^3}\right)\left(\frac{\mathcal{K}}{L^2}-\frac{1}{r^2}\right)-\frac{1}{\sigma^2}.
\end{equation} 

The effective potential $V_{\rm eff}$ defined in the previous equation will help us to determine the existence of geodesics anchored in the boundary. \\
The quBTZ black hole is asymptotically AdS and therefore we have $\lim_{r\rightarrow \infty}V_{\rm eff}\sim -r^2\mathcal{K}$, then the potential is positive at infinity when $\mathcal{K}=-1$. This implies that there are no time-like geodesics connecting two points of the boundary. This fact applies to all asymptotically AdS spacetimes. Therefore, we will be concerned only about null and space-like geodesics connecting points in the boundary.

\subsection{Null geodesics}
We start by studying null geodesics and we set all constant to one\footnote{In Appendix \ref{App:3} we discuss the general case.}. We want to determine if there are null geodesics of $Type\,\mathcal{C}$ that causally connect two given points in the boundary. We set $\mathcal{K}=0$, and write the effective potential \eqref{pot:fin}  as:

\begin{equation}\label{potn:1}
V_{\rm eff}(r)=1+\frac{4\kappa x_{1}^2}{(3-\kappa x_{1}^2)^2r^2}-\frac{8(1-\kappa x_{1}^2)}{(3-\kappa x_{1}^2)^3 r^3}-\frac{1}{\sigma^2}.
\end{equation}

Due to the fact that the mass of the quBTZ black hole depends on $\kappa x_{1}^2$ and not on $\kappa$ and $x_{1}^2$  separately, we are going to study three different cases\footnote{Note that the case $\kappa=0$ is equivalent to the case $x_1=0$ and therefore the three cases that we are going to study belong to Branch 1.} $\kappa x_{1}^{2}=0,\pm 1$.

\begin{figure}[h!]
\includegraphics[scale=0.62]{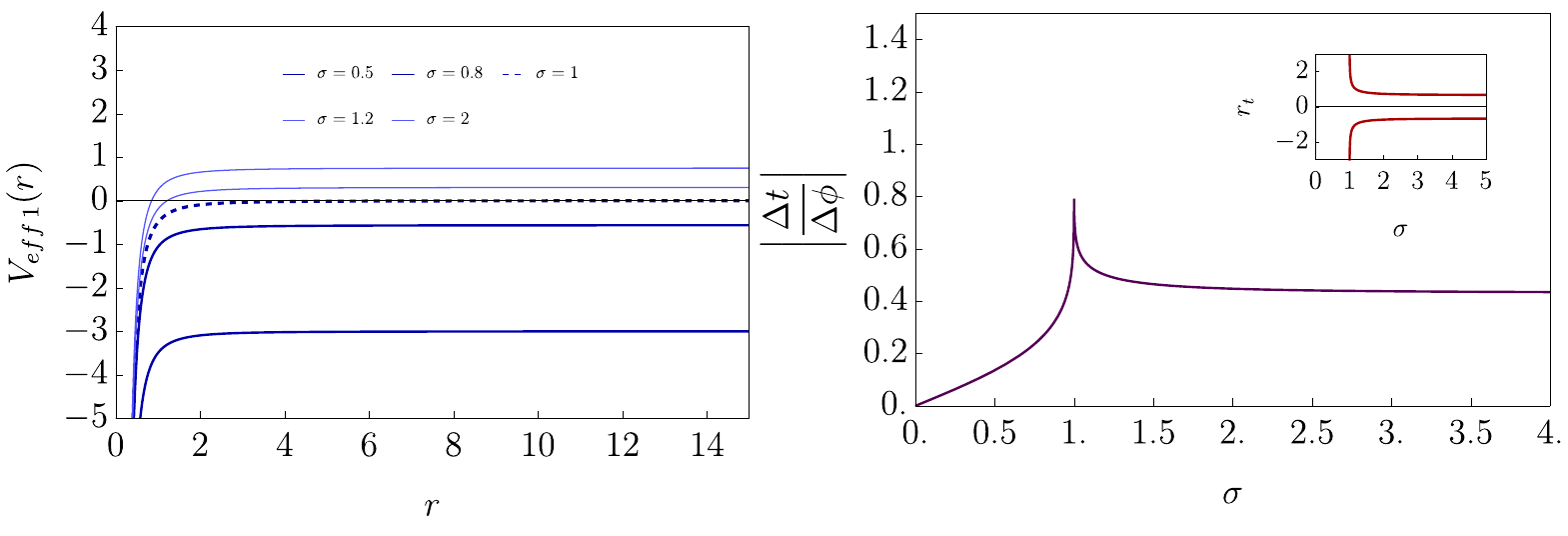}
\caption{In the left panel we have plotted the effective potential $V_{\rm eff\,1}$ as a function of the radial coordinate and for different values of the impact parameter $\sigma$. The light blue curves represent the $V_{\rm eff\,1}$ for $\sigma=1.2$ and $\sigma=2$. The dashed line corresponds to the value $\sigma=1$  and the dark blue curves correspond to the values $\sigma=0.5$ and $0.8$. In the right panel we have plotted $\left\vert\frac{\Delta t}{\Delta \phi}\right\vert$ as a function of $\sigma$. The peak of the function is at $\sigma=1$. In any case there is never a point where it reaches $1$. The inset plot in the right panel correspond to the return point $r_t$  as a function of $\sigma$.}
\label{nullfin}
\end{figure}

\textbf{Case $\kappa x_{1}^{2}=0$}. Now the potential \eqref{potn:1} becomes
\begin{equation}
V_{\rm eff\,1}(r)=1-\frac{1}{\sigma^2}-\frac{8}{27 r^3}.
\end{equation}
When $r\rightarrow\infty $ then  $V_{\rm eff\,1}(r)=1-\frac{1}{\sigma^2}$ which is negative if $\sigma^2<1$, in other words, when $\sigma^2 \in (0,1]$ there are null geodesics departing  from the boundary of  the quBTZ black hole. The behavior of $V_{\rm eff\,1}$ for large values of $r$ can be clearly seen. Only the potentials with $\sigma^2<1$ will lead to geodesics with at least one end attached to the boundary. The roots of $V_{\rm eff\,1}(r)$ are going to be
\begin{equation}\label{retp:1}
r_{t}=\pm\frac{2\sigma^{2/3}}{3(\sigma^2-1)^{1/3}},
\end{equation}
hence, the condition $\sigma^2<1$  that brings the potential to be negative at infinity now leads to a negative/positive turning point. If there is no horizon or singularity then geodesics can go to negative values of $r$ but then return to the boundary.  \\
The potential $V_{\rm eff\,1}$ is plotted in the left panel of Fig. \ref{nullfin}. There are two zones clearly divided by $\sigma_c=1$. The values above $\sigma_c$ (light blue) lead to a effective potential that is positive at infinity, therefore it is not possible to have geodesics anchored in the boundary. The blue dashed line represents the value $\sigma_c=1$, this leads to a potential that vanishes at infinity and therefore it represents geodesics living in the boundary, namely geodesics of $Type\,\mathcal{A}$. The dark blue line corresponds to values $\sigma>\sigma_c$ and in all these cases the potential becomes negative at infinity, therefore there are geodesics anchored in the boundary. Due to the fact that the potential curves with $\sigma^2<1$ do not have roots, there are no turning points. In the plot inside the right panel, $r_t$ is plotted as a function of $\sigma$ (dark red curve).  There  exists a turning point only for  $\sigma>1$, therefore the only possibilites are geodesics of $Type\,\mathcal{A}$ and $Type\,\mathcal{B}$. In the right panel of Fig. \ref{nullfin} we have plotted (purple curve)  $|\frac{\Delta t}{\Delta \phi}|$ as a function of $\sigma$. We have that $|\frac{\Delta t}{\Delta \phi}|<1$ for all values of $\sigma$, therefore for every two points in the boundary, the distance between them is going to be light-like. Thus, geodesics lying in the boundary will be of the null type. 

In order to  study causality we use equations \eqref{dt} and \eqref{dp} with $f(r)$ defined in \eqref{f:2}, thus

\begin{eqnarray}
\frac{dt}{dr}&=&\frac{r}{\left(r^2-\frac{8}{27r}\right)\sqrt{r^2-\sigma^2\left(r^2-\frac{8}{27r}\right)}},\\
\frac{d\phi}{dr}&=&\frac{\sigma}{r\sqrt{r^2-\sigma^2\left(r^2-\frac{8}{27r}\right)}},
\end{eqnarray}

whose respective solutions are 
\begin{eqnarray}
t(r)&=&-\frac{81\sqrt{3}r^{7/2}AF_1\left(\frac{7}{6},\frac{1}{2},1,\frac{13}{6},\frac{27r^3(\sigma^2-1)}{8\sigma^2},\frac{27r^3}{8}\right)}{56\sqrt{2}\sigma}+C_1,\\
\phi(r)&=&\frac{3\sqrt{3r}}{\sqrt{2}}H_2F_1\left(\frac{1}{6},\frac{1}{2},\frac{7}{6},\frac{27r^3(\sigma^2-1)}{8\sigma^2}\right)+C_2,
\end{eqnarray}
where $C_1$ and $C_2$ are integration constants, $AF_1(a,b^1,b^2,c,x,y)$ is the Appell hypergeometric function of two variables $x,y$ and $H_2F_1(a,b,c,z)$ is the hypergeometric function. We calculate $\Delta t(r_t)$ and $\Delta \phi(r_t)$ such that $r_t$ is given by \eqref{retp:1}, 
therefore
\begin{equation}
\Bigg|\frac{\Delta t}{\Delta \phi} \Bigg| =\frac{\sigma H_2F_1\left(1,\frac{7}{6},\frac{5}{3},\frac{\sigma^2}{\sigma^2-1}\right)}{4(\sigma^2-1)}, \,\,\sigma\in (0,1].
\end{equation}

When $\sigma\rightarrow 1$ then $\lvert \frac{\Delta t}{\Delta \phi}\rvert\rightarrow 1$, showing that when $E=L$ there is a light-like separation between points in the boundary. Moreover, when $\sigma \rightarrow 1$  then $r_t\rightarrow \infty$ leading to the conclusion that in this limit the geodesics become of the $Type\,\, \mathcal{A}$.\\
\textbf{Case $\kappa x_{1}^2=1$}. The effective potential \eqref{potn:1} becomes
\begin{equation}\label{pot:2}
V_{\rm eff\,2}(r)=1+\frac{1}{r^2}-\frac{1}{\sigma^2}, 
\end{equation}
which is the potential of a pure AdS spacetime with spherical symmetry. This potential is monotonically decreasing towards the value $1-\frac{1}{\sigma^2}$, therefore, if we want a null geodesic departing from the boundary we should impose $\sigma^2<1$. For a fixed $\sigma$ the potential \eqref{pot:2} has a root at $r_t=\sigma^2/(\sqrt{1-\sigma^2})$, then as before, $\sigma^2<1$ implies the existence of null geodesics of $Type\, \mathcal{C}$. The solution of equations \eqref{dt} and \eqref{dp} are \cite{He:2024emd}
\begin{eqnarray}
t(r)&=&\arctan\left(\sqrt{(1-\sigma^2)r^2-\sigma^2}\right)-\frac{\pi}{2}\\
\phi(r)&=&\arctan\left(\frac{\sigma}{\sqrt{(1-\sigma^2)r^2-\sigma^2}}\right)-\frac{\pi}{2}
\end{eqnarray}
The length of a geodesic anchored in the boundary can be calculated then $|\frac{\Delta t}{\Delta \phi}|=1.$
The previous result implies that the endpoints of a null geodesic if $Type \,\mathcal{C}$ have always a light-like separation. As $\sigma^2 \rightarrow 1$ the turning point $r_t$ tends to infinity, which is the exact location of the boundary, therefore, in this case the geodesic changes to that of $Type \, \mathcal{A}$, for a detailed study see \cite{He:2024emd}.\\
\textbf{Case $\kappa x_{1}^2=-1$}. In this final case the effective potential becomes: 
\begin{equation}\label{pot:3}
V_{\rm eff\,3}(r)=1-\frac{1}{4\,r^2}-\frac{1}{4\,r^3}-\frac{1}{\sigma^2}.
\end{equation}

As in previous cases, the condition $\sigma^2<1$ is needed for having a negative effective potential at infinity. On the other hand, the roots of $V_{\rm eff\,3}(r)$ are given by $r_t=\pm\sigma/{\sqrt{\sigma^2-1}}$, therefore, in order to have a positive real root that corresponds to a turning point it is required that $\sigma^2>1$. It implies that there are no null geodesics of the $Type\,\, \mathcal{C}$. There are null geodesics departing from the boundary at infinity, however, there is no turning point and therefore the geodesic does not return to the boundary. \\
The three cases studied belong to Branch 1, however in order to study an example belonging to Branch 2 we set as an example $-\kappa x_1^2=4$, then the effective potential becomes 
\begin{equation}\label{pot:4}
V_{\rm eff\,4}(r)=1-\frac{16}{49\, r^2}-\frac{40}{343\,r^3}-\frac{1}{\sigma^2},
\end{equation}
which is negative at $r=+\infty$ when $\sigma^2\leq1$, therefore, there are geodesics with an endpoint at the boundary. However,  the potential defined in \eqref{pot:4} does not have positive real roots for $\sigma^2\leq1$, certainly not a turning point, then all geodesics with an end at the boundary are of $Type\,\,\mathcal{B}.$

\subsection{Space-like geodesics}

In this section we are going to perform a similar analysis to the one already presented for null geodesics, we now consider spacelike geodesics by setting $\mathcal{K}=1$. The effective potential \eqref{geor} now becomes

\begin{equation}\label{pot:5}
V_{\rm eff}(r)=1+\frac{4\kappa x_{1}^2}{(3-\kappa x_{1}^2)^2r^2}-\frac{8(1-\kappa x_{1}^2)}{(3-\kappa x_{1}^2)^3 r^3}-\frac{1}{\sigma^2}-\frac{1}{L^2}\left(\frac{8(1-kx_1^2)}{(3-\kappa x_1^2)^3 r}-\frac{4\kappa x_1^2}{(3-\kappa x_1^2)^2}\right)-\frac{r^2}{L^2}.
\end{equation}
\textbf{Case $kx_1^2=0$ }. The potential \eqref{pot:5} transforms to
\begin{equation}\label{pots:1}
V_{\rm eff\,5}=1-\frac{1}{\sigma^2}-\frac{8}{27 r^3}+\frac{8}{27 L^2 r}-\frac{r^2}{L^2},
\end{equation}

When $r\rightarrow \infty$ then $V_{\rm eff\,5}\rightarrow -\infty$, the potential decreases towards negative values when the radial coordinate grows, this leads to the existence of geodesics with an end at the boundary. 

\begin{figure}[h]
\centering
\includegraphics[scale=0.7]{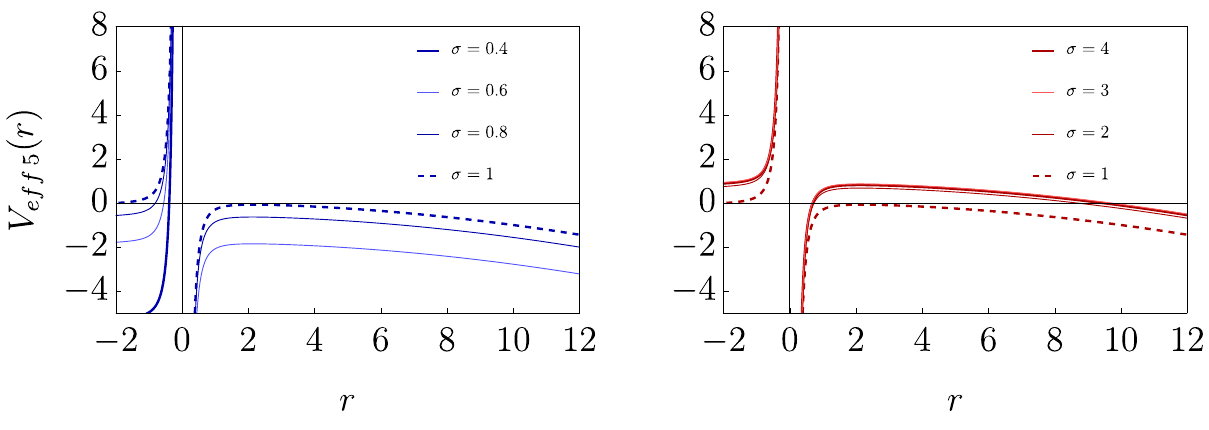}
\caption{We have plotted the potential $V_{\rm eff\,5}$ for different values of $\sigma$. We have set $\mathcal{K}=1$ and fixed the value of the momenta to $L=10$, therefore each different value of $\sigma$ corresponds to different value of the energy $E$. We have also set $kx^2=0$. In the left panel we have plotted the potential $V_{\rm eff\,5}$ for $\sigma$ values lower than one. In the right panel the potential is plotted for $\sigma$ values higher than one. }
\label{sppot:1}
\end{figure}

In order to identify a turning point we want to find the roots of  $V_{\rm eff\,5}$, then we have to determine the roots of the following equation (see Appendix \ref{App:1} for an analytical approach and restrictions over parameter $L$):
\begin{equation}\label{pol:2}
r^5-r^3 \left(1-\frac{1}{\sigma^2}\right)L^2-\frac{8}{27}r^2+\frac{8}{27}L^2=0.
\end{equation}
If $\sigma^2>1$ then the polynomial \eqref{pol:2} has two changes of signs therefore, according to Descartes's rule of signs, it has two positive roots at most or it has zero roots and therefore no turning point. On the other hand, if $\sigma^2<1$ we have that the potential \eqref{pol:2} has two changes of signs and therefore the effective potential  has either two positive roots or none. We can conclude that for any value of $\sigma^2$ the effective potential $V_{\rm eff\,5}$ has either two positive roots or none. In Fig. \ref{sppot:1} we have plotted the potential \eqref{pots:1}. In both panels the value of momenta is fixed to $L=10$ and we have set $\kappa x_1^2=0$. From both panels it can be easily seen that for large values of  $r$ the potential is negative, therefore there should be geodesics with an endpoint  at the boundary. However, only for some values $\sigma_c >1$ there are points where the  potential $V_{\rm eff\,5}$ vanishes, leading to the existence of turning points and therefore of geodesics of $Type\,\,\mathcal{C}$. 
A detailed analytic study of the allowed parameter ranges in $L$ and of the existence of real positive turning points is presented in Appendix \ref{App:1}.

\begin{figure}[h]
\centering
\includegraphics[scale=0.52]{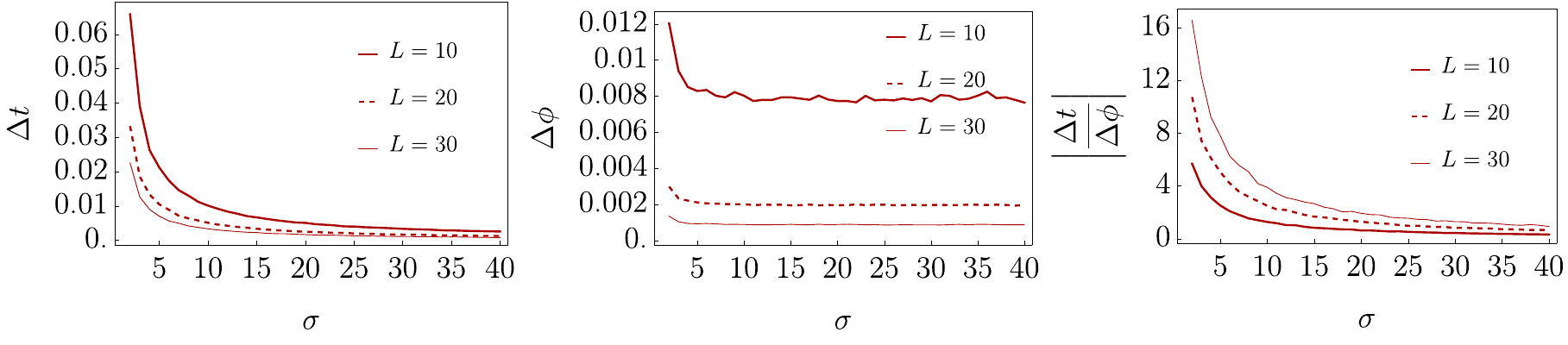}
\caption{The plots show results found by numerical integration. We have set $\kappa x_1^2=0$. In the left panel we have plotted $\Delta t$ defined in \eqref{dt}. The lower integration limit is calculated also numerically, which corresponds to the highest root of $V_{\rm eff\,5}$. Every point corresponds to $\sigma=1,\dots,20$. We have joined all points with a curve. Each curve  corresponds to the values of $L=10,L=20,L=30$. In the center panel we plotted $\Delta \phi$ defined in \eqref{dp}. The parameters $\sigma$ and $L$ are the same in the three panels. In the right panel we plot the quotient $\Delta t/\Delta \phi$}.
\label{spcausality:1}
\end{figure}

Let's see if the points at the boundary are causality connected. We use equations \eqref{dt} and \eqref{dp} with $f(r)$ defined in \eqref{f:2} with $\kappa x_1^2=0$. Unfortunately, the integrals can not be done analytically and we rely on a numeric integration method. In Fig. \ref{spcausality:1} we have plotted the curves  $\Delta t$ and $\Delta \phi$. We have set $L=10, L=20, L=30$ and $\sigma=1,\dots,20$. Due to the fact that $\sigma=L/E$  every value of $\sigma$ corresponds to a value of $E$ with $L$ fixed. In the right panel we have the plots of the curves of the quotient $\Delta t/\Delta \phi$. For small values of $\sigma$ the quotient is really high, meaning that the separation between points at the boundary is time-like, as $\sigma$ grows the value of the quotient reduces. For high values of $\sigma$ and $L$ the curve crosses the null limit $\Delta t=\Delta \phi$, and then the distance between the points in the boundary becomes spacelike.

\textbf{Case $kx_1^2=1$ }. The potential \eqref{pot:5} transforms to
\begin{equation}
V_{\rm eff\,6}=1-\frac{1}{\sigma^2}+\frac{1}{r^2}-\frac{1+r^2}{L^2}.
\end{equation}

The effective potential $V_{\rm eff\,6}$ is negative when $r\rightarrow +\infty$, and it goes to $+\infty$ when $r\rightarrow 0$, it is monotonically decreasing in the interval $[0,\infty)$ and therefore the existence of geodesics with one end at the boundary is guaranteed. In order to determine the existence of turning points we have to find the solutions of the equation
\begin{equation}\label{pol:6}
\frac{r^4}{L^2}-r^2\left(1-\frac{1}{\sigma^2}-\frac{1}{L^2}\right)-1=0.
\end{equation}
 
The polynomial \eqref{pol:6} has  one change of sign for all possible values of $\sigma^2$ and $L^2$, therefore the effective potential $V_{\rm eff\,6}$ will have either one positive root or none\footnote{When $\sigma^2>L^2/(L^2-1)$ for $L\neq 1$ or when $\sigma^2>1$ for $L=1$ the polynomial \eqref{pol:6} will have only one change of sign, therefore, by the Descartes's rule of signs, it will have one positive root or none, however in any other case there will be  only one change of sign also, therefore by the Descartes's rule of signs the polynomial \eqref{pol:6} has either one positive real root or none. We conclude that independently of the values of the parameters $\sigma^2$ and $L^2$ the effective potential $V_{\rm eff\,6}$ will have either only one positive real root or none.}.

\begin{figure}
\centering
\includegraphics[scale=0.6]{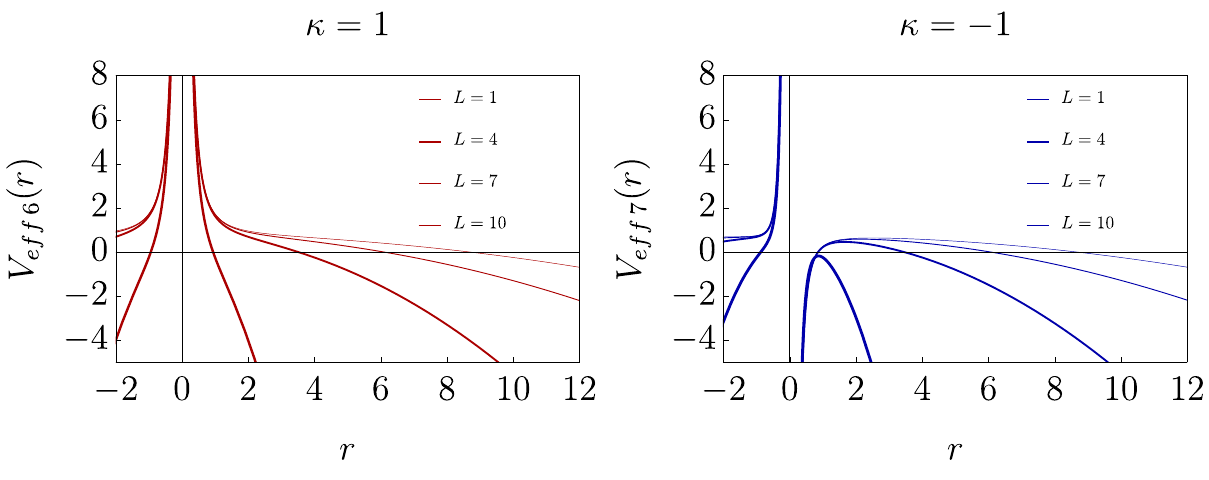}
\caption{Potentials $V_{\rm eff\,6}$ and $V_{\rm eff\,7}$ for $L=1,4,7,10$ and $\sigma=2$. The left panel (red curves) corresponds to the effective potential $V_{\rm eff\,6}$ with $\kappa x_1^2=-1$. In the right panel the effective potential $V_{\rm eff\,7}$ is plotted (blue curves) with the same parameters as $V_{\rm eff\,6}$  but with $\kappa x_1^2=1$.}
\label{pot67}
\end{figure}

In Fig. \ref{pot67} we have plotted the effective potential $V_{\rm eff\,6}$ in the left side panel (red curves). For values of $L$ higher then one the effective potential  $V_{\rm eff\,6}$   will have one root, which is going to be the turning point. The potential  is negative for larger values of $r$. Let us show how is the distance between points in the boundary behaves.

\begin{figure}[h]
\centering
\includegraphics[scale=0.52]{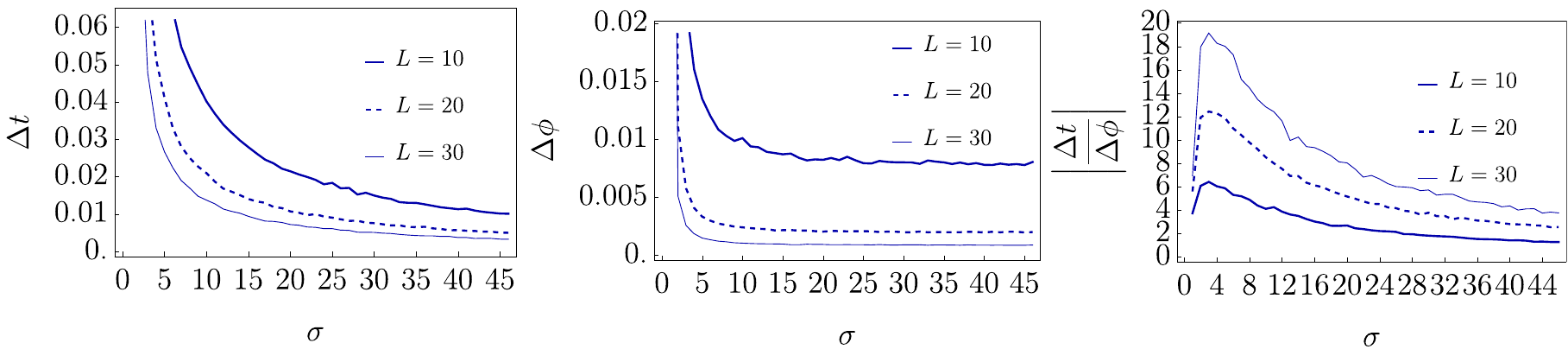}
\caption{The plots show results found by numerical integration. We have set $\kappa x_1^2=1$.In the left panel we have plotted $\Delta t$ defined in \eqref{dt}. The lower integration limit is calculated also numerically, which corresponds to the highest root of $V_{\rm eff\,5}$. Every point corresponds to $\sigma=1,\dots,20$. We have joined all points with a curve. Each curve  corresponds to the values of $L=10,L=20,L=30$. In the center panel we plotted $\Delta \phi$ defined in \eqref{dp}. The parameters $\sigma$ and $L$ are the same three panels. In the right panel we plot the quotient $\Delta t/\Delta \phi$}
\label{spcausality:3}
\end{figure}

\begin{figure}[h]
\centering
\includegraphics[scale=0.52]{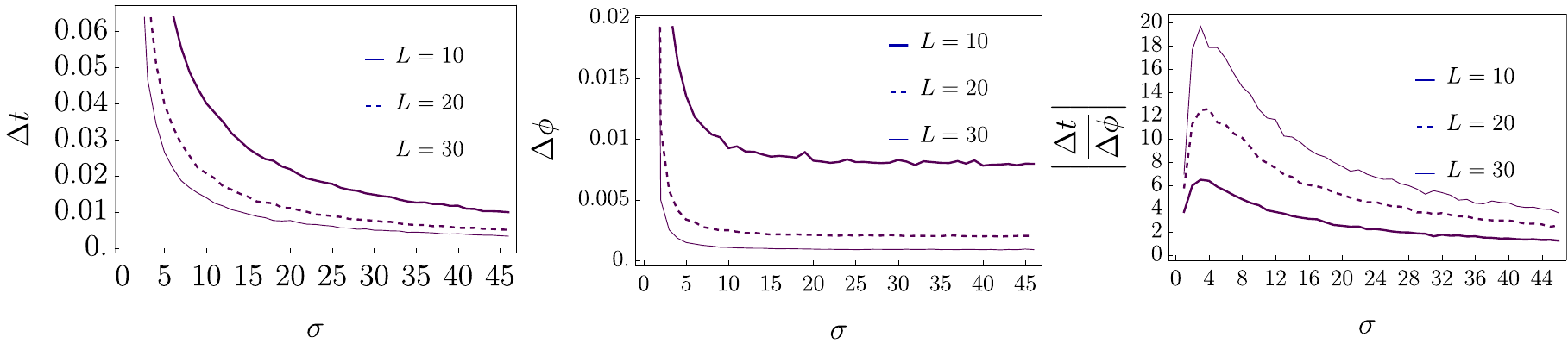}
\caption{The plots show results found by numerical integration. We have set $\kappa x_1^2=-1$.In the left panel we have plotted $\Delta t$ defined in \eqref{dt}. The lower integration limit is calculated also numerically, which corresponds to the highest root of $V_{\rm eff\,5}$. Every point corresponds to $\sigma=1,\dots,20$. We have joined all points with a curve. Each curve  corresponds to the values of $L=10,L=20,L=30$. In the center panel we plotted $\Delta \phi$ defined in \eqref{dp}. The parameters $\sigma$ and $L$ are the same three panels. In the right panel we plot the quotient $\Delta t/\Delta \phi$.}
\label{spcausality:2}
\end{figure}

In Fig. \ref{spcausality:3} we have plotted the results. The curves are joining the values obtained from the integration of \eqref{dt} and \eqref{dp}. Each value in the curve corresponds to the same value of $L$ with $\sigma=1,\dots,20$. The quotient $\Delta t/ \Delta \phi$ is plotted in the right panel of Fig. \ref{spcausality:2}. The values of that quotient are above one for small values of $\sigma$ and $L$ which implies that the separation between points at the boundary is time-like, for big values of $\sigma$  all curves eventually cross the null limit $\Delta t=\Delta \phi$ and therefore the separation between points at the boundary connected by a geodesic become space-like.

\textbf{Case $\kappa x_1^2=-1$}. The effective potential can be written
\begin{equation}
V_{\rm eff\,7}=1-\frac{1}{\sigma^2}+\frac{1}{4L^2}-\frac{r^2}{L^2}+\frac{1}{4 L^2 r}-\frac{1}{4 r^3}-\frac{1}{4 r^2}
\end{equation}

When $r \rightarrow \infty$ the potential behaves as $V_{\rm {eff\,7}}\rightarrow -\infty$. Moreover, when $r\rightarrow 0$ then $V_{\rm {eff\,7}}$ diverges. In order to find the turning points we have to study the solutions of the equation
\begin{equation}
\frac{r^5}{L^2}-r^3\left(1-\frac{1}{4L^2}-\frac{1}{\sigma^2}\right)-\frac{r^2}{4L^2}+\frac{r}{4}+\frac{1}{4}=0.
\end{equation}  

The left side of the previous equation, independently of the values taken by the parameters $\sigma$ and $L$ have two changes of sign and therefore the effective potential $V_{\rm eff\,7}$ has either two real positive roots or none.\\
In Fig. \ref{pot67}, in the right panel we have plotted the potential $V_{\rm eff\,7}$. For all values of $L$ the potential goes to $-\infty$. Note that we have fixed $\sigma=2$ for both panels. In the graph we can see that the potential $V_{\rm eff\,7}$ has zero roots for $L=1$, however when $L$ increases the potential has two roots in accordance with our analysis. The root with the maximum value is going to be the turning point of the geodesics. Now let us analyze the causality between points at the boundary. The integrals \eqref{dt} and \eqref{dp} have to be evaluated numerically as the roots of the effective potential $V_{\rm eff\,7}$. 

\begin{figure}[h!]
\centering
\includegraphics[scale=0.6]{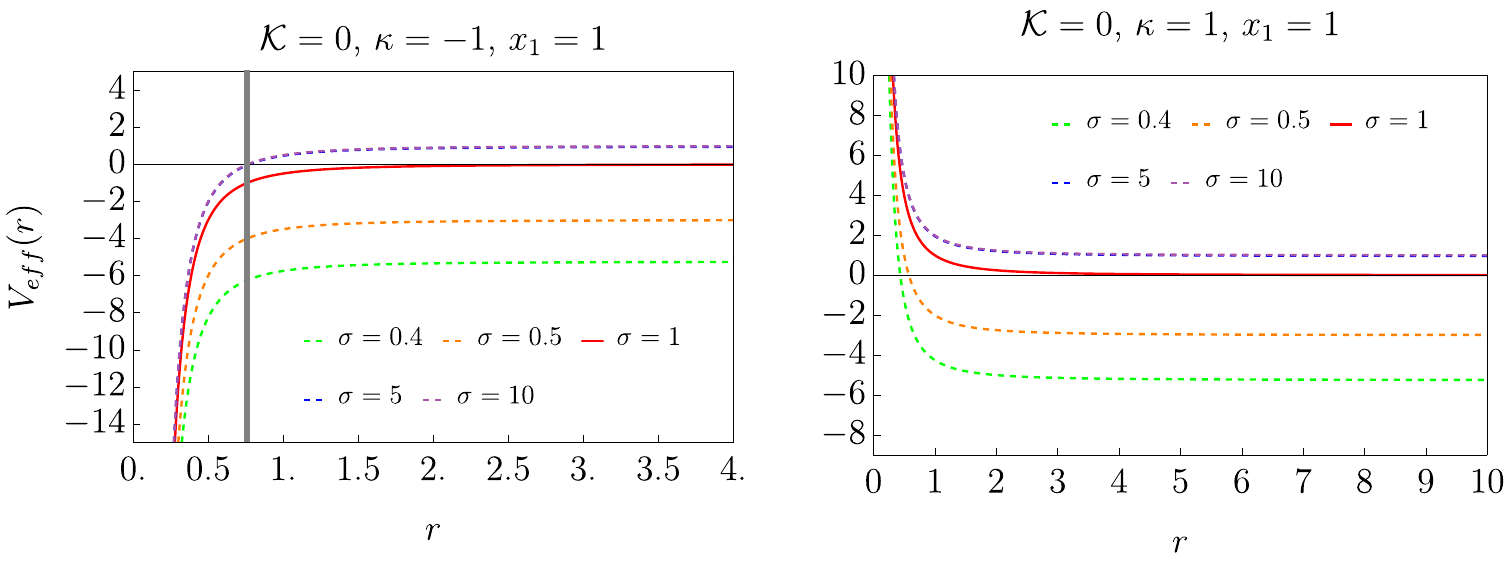}
\caption{In the left panel the effective potential for the uncharged case $(q=0)$, with $\mathcal{K}=0$, $\kappa=-1$, $x_1=1$, is plotted. The dashed curves represent the effective potential for different values of $\sigma$, while the continuous red curve represents the effective potential for $\sigma=1$. The vertical gray line stands for the location of the horizon of the black hole. In the right panel the effective potential for  $\mathcal{K}=0$, $\kappa=1$, $x_1=1$ and $q=0$ is plotted, there is no horizon for this case. We have set $L=20$ in all cases.} 
\label{pothor1}
\end{figure}

 In Fig. \ref{spcausality:2} we have plotted the results of the numerical integration for $\Delta t$, $\Delta \phi$  and the quotient $\Delta t/\Delta \phi$. Every curve is built by joining the points obtained after numerical integration by setting a fixed $L$ and $\sigma=1,\dots,20$. While the values  of $\Delta \phi$ are decreasing for all $L$ and increasing values of $\sigma$, the values of $\Delta t$ are decreasing faster, it implies that at some point the null limit will be reached and surpassed. This can be inferred from the right panel of Fig. \ref{spcausality:2}. The values of the quotient $\Delta t/\Delta \phi$ are higher the one for small values of $\sigma$, leading to a timelike separation in the boundary, however the quotient decreases until reaching the null limit when $\sigma$ increases. Thus, the distance between points at the boundary will become null and at, some sufficiently high value of $\sigma$, it will become spacelike.\\
The presence of horizons can be determinant when the existence of turning points is studied. If a geodesic with one end at the boundary crosses the horizon then it is impossible that it returns again to the boundary. Therefore, if there is a horizon the turning point should be outside that horizon. In Fig. \ref{pothor1} we have plotted the effective potential for the uncharged case $(q=0)$, with $\mathcal{K}=0$, $\kappa=-1$, $x_1=1$ (left panel). The horizon is located at $r_{H}=0.76069$. The potentials that become negative at infinity (orange and green curves) do not have turning points, then there are no geodesics of the $Type\,\,\mathcal{C}$. In this case the presence of horizon does not change the fact that there are not geodesics of  $Type\,\,\mathcal{C}$. In the right panel the behaviour of the effective potential is different. It has roots, and therefore, turning points. Moreover, there are values of $\sigma$ such that the potential becomes negative at infinity. However, there is not a horizon located in the vecinity of the roots, therefore there are geodesics of $Type\,\,\mathcal{C}$.\\

\begin{figure}[h]
\centering
\includegraphics[scale=0.6]{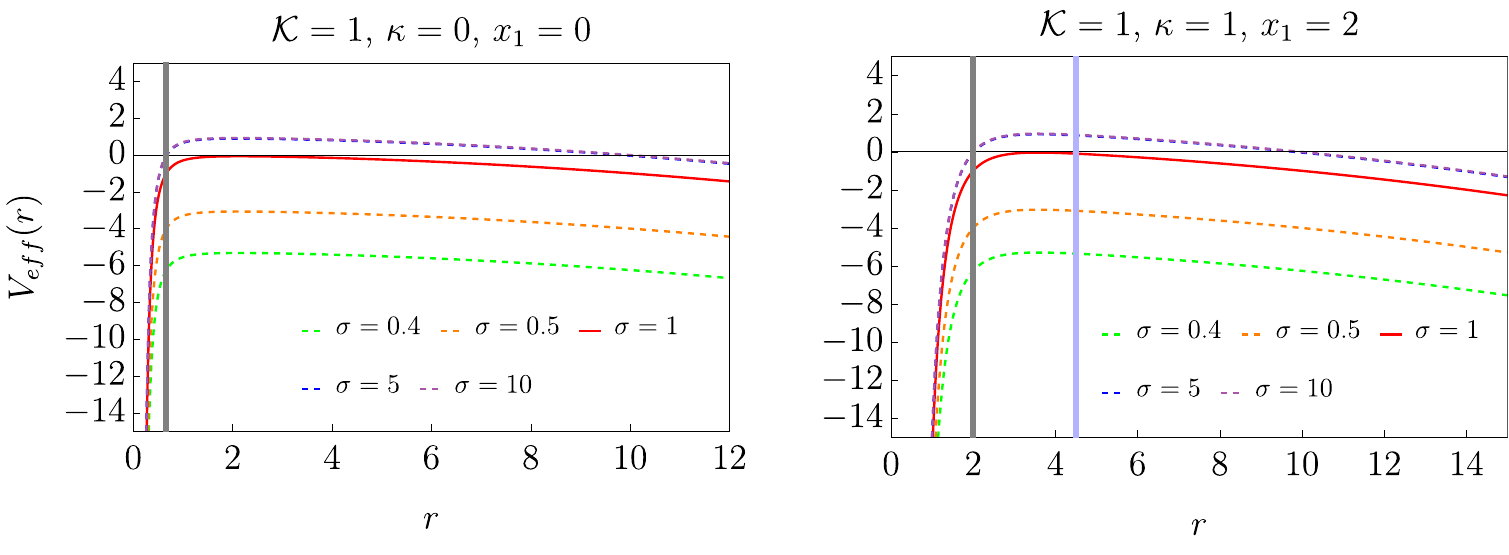}
\caption{In the left panel the effective potential for the uncharged case $(q=0)$, with $\mathcal{K}=1$, $\kappa=-1$, $x_1=1$, is plotted. The dashed curves represent the effective potential for differnet values of $\sigma$, while the continuous red curve represents the effective potential for $\sigma=1$. The vertical gray line stands for the location of the horizon of the black hole. In the right panel the effective potential for  $\mathcal{K}=1$, $\kappa=1$, $x_1=1$ and $q=0$ is plotted. The vertical gray line stands for the location of the horizon of the black hole  and the vertical blue line signals the location of the photon sphere. We have set $L=20$ in all cases.} 
\label{pothor2}
\end{figure}

In Fig. \ref{pothor2} we have plotted the effective potential for different values of $\sigma$ and with $\mathcal{K}=1$, $\kappa=0$, $x_1=0$. In the left panel can see that the curve for $\sigma=10$ vanishes at two points, one point that is inside the horizon but the other (the bigger one) is outside the horizon. Moreover, the potential becomes negative at infinity therefore we can safely argue that there are geodesics of $Type\,\,\mathcal{C}.$ Simlarly, the potential with $\sigma=1$ has one root outside of the horizon and becomes negative at infinity ensuring the existence of geodesics of  $Type\,\,\mathcal{C}$. In the right panel we have a similar plot for $\mathcal{K}=1$, $\kappa=1$, $x_1=2$. The black hole has a photon sphere which is representd by the blue vertical line. All curves become negative at infinity, however not all of them have roots. The purple dashed curve ($\sigma=0.4$) and the red continuous curve ($\sigma=1$) have two two and one root repectively, then in both cases, there are geodesics of $Type\,\,\mathcal{C}$.  The existence of a horizon or a photon sphere could be used for determining the existence of geodesics of   $Type\,\,\mathcal{C}$. In the next section we are going to show that the presence of an electrical charge $q$ in the potential can change the behavior of the geodesics drastically. 


\section{The charged quantum BTZ black hole}\label{sec:4}
The quantum BTZ black hole  studied in the previous section can be generalized to include electrical charge. The blackening factor of the metric \eqref{qBTZ:1} now includes an electrical charge in the following way \cite{Feng:2024uia,Kinnersley:1970zw}:

\begin{equation}
f(r)=\frac{r^2}{\ell_{3}^2}-8\mathcal{G}_3M-\frac{\ell F(M)}{r}+\frac{\Delta^4 q^2 \ell^2}{r^2}
\end{equation}
where
\begin{equation}
\Delta=\frac{2x_1}{3-\kappa x_1^2+q^2x_1^4},
\end{equation}
and 
\begin{equation}
F(M)=\frac{8(1-\kappa x_1^2-q^2x_1^4)}{(3-\kappa x_1^2+q^2x_1^4)^3}.
\end{equation}
An outer horizon $r_{+}$ and an inner Cauchy horizon $r_{-}$ can be found by solving $f(r_{\pm})=0.$ The mass of the black hole is given by
\begin{equation}
M=-\frac{1}{2G_3}\frac{\ell}{\ell_4}\frac{\kappa x_1^2}{(3-\kappa x_1^2+q^2x_1^4)^2}.
\end{equation}

When $\kappa=+1$ then $x_1\in (0,1)$ then 

\begin{equation}\label{cqbtzcond}
0\leq q\leq \sqrt{\frac{1-\kappa x_1^2}{x_1^4}},
\end{equation}
when $q=0$ we have $x_1=1$ . On the other side, when $\kappa=0,-1$ then $x_1\in (0,\infty)$.

The effective potential now becomes
\begin{eqnarray}\label{potcharg:1}
V_{\rm eff}(r)&=&1-\frac{\mathcal{K}r^2}{L^2}-\frac{1}{\sigma^2}-\frac{8\mathcal{K}(1-\kappa x_1^2-q^2 x_1^4)}{L^2 (3-\kappa x_1^2+q^2 x_1^4)^3r}+\frac{4 x_{1}\left(\kappa(3-\kappa x_1^2+q^2x_1^4)^2 L^2-4 \mathcal{K}q^2 x_1^3\right)}{L^2(3-\kappa x_{1}^2+q^2x_1^4)^4r^2}\nonumber\\
& &-\frac{8(1-\kappa x_{1}^2-q^2x_1^4)}{(3-\kappa x_{1}^2+q^2x_1^4)^3 r^3}+\frac{16x_1^4 q^2}{(3-\kappa x_1^2+q^2x_1^4)^4 r^4}.
\end{eqnarray}

\begin{figure}[h]
\centering
\includegraphics[scale=0.7]{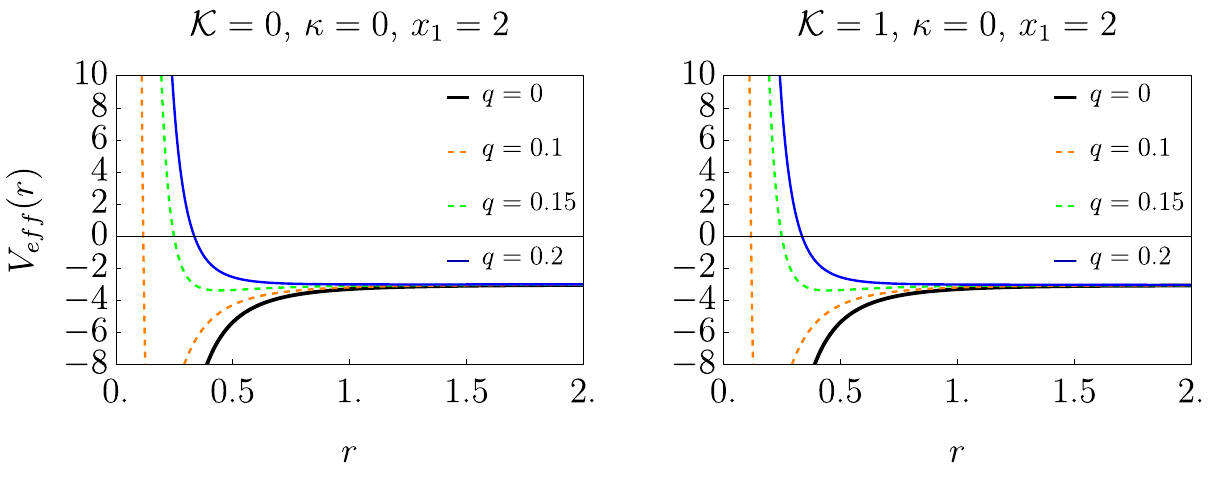}
\caption{In the left panel we have plotted the  potential \eqref{potcharg:2}, for the null case ($\mathcal{K}=0$ ). The black continuous curve represents the potential for the uncharged black hole $q=0$ and the blue continuous curve represents the potential with $q=0.2$. The dashed lines represent the potential with $q=0.1,0.15$. In the right panel we have a similar plot of the spacelike potential ($\mathcal{K}=1)$ for the same values of the charge $q$. We have set $\kappa=0$ and $x_1=2$ and $L=10$ in both panels.}
\label{potc:1}
\end{figure}

When $r\rightarrow \infty $ the effective potential \eqref{potcharg:1} behaves as $\lim_{r\rightarrow \infty} V_{\rm eff}(r)\sim -\mathcal{K}r^2$, thus when $\mathcal{K}=-1$ the effective potential at infinity becomes positive, therefore there is not timelike geodesics anchored in the boundary.  We are going to study the existence of null and spacelike geodesics with ends at infinity. \\
Let us analyze the null case $\mathcal{K}=0$, then the effective potential \eqref{potcharg:1} becomes 

\begin{equation}\label{potcharg:2}
V_{\rm eff}=1-\frac{1}{\sigma ^2}+\frac{16 q^2 x_1^4}{r^4 \left(3-\kappa x_1^2+q^2 x_1^4\right)^4}-\frac{8 \left(1-\kappa x_1^2-q^2
   x_1^4\right)}{r^3 \left(3-\kappa x_1^2+q^2 x_1^4\right)^3}+\frac{4 \kappa x_1}{r^2 \left(3-\kappa
   x_1^2+q^2 x_1^4\right)^2}
   \end{equation}
The potential \eqref{potcharg:2} will become negative at infinity when $\sigma^2<1$. In Fig. \ref{potc:1} we have plotted the potential \eqref{potcharg:2}  with electrical charge $q=0, 0.1, 0.5, 1$ and $\kappa=1$ and $x_1=1$.

 The potential with $q=0$ is the black continuous curve. The potential is negative at infinity and the existence of a turning point, a root of the potential, is also guaranteed. The condition \eqref{cqbtzcond} implies that $q=0$ when $x_1$, however we have plotted the potential \eqref{potcharg:2} for values that do not satisfy the condition. In this case, the introduction of the electrical charge does not change the global behavior of the potential. From this we can conclude that there are null geodesics of $Type\, \mathcal{C}$.

\begin{figure}[h]
\centering
\includegraphics[scale=0.7]{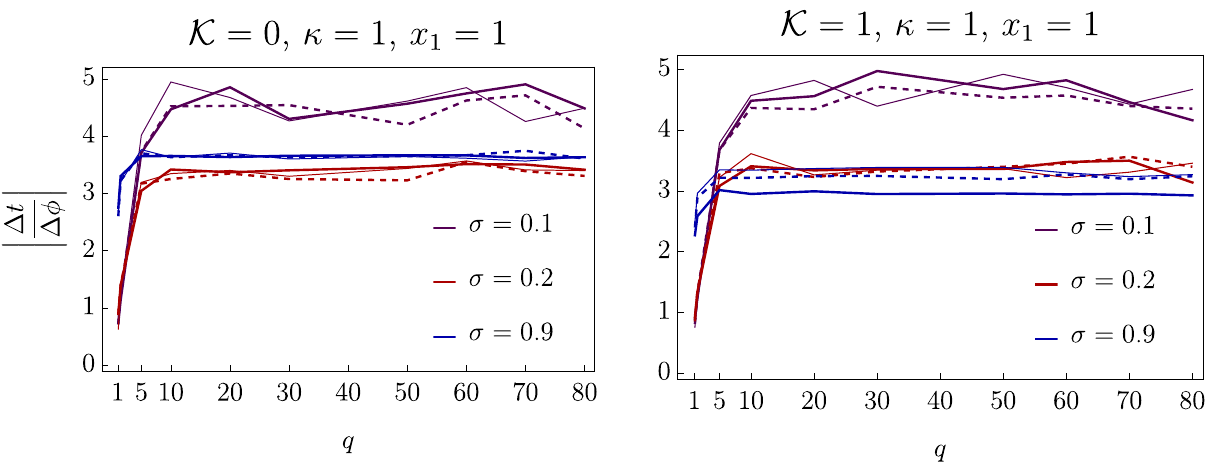}
\caption{The plots show the quotient $|\Delta t|/|\Delta\phi|$ for different values of $\sigma$ and $L$. The purple curves represent the quotient for $\sigma=0.1$ and each purple curve is found by joining the points for every value of the electrical charge $q=1.1, 1.5, 5, 10, 20, 30, 50, 60, 70, 80$. The three curves represent three different values of momenta,  $L=10$ (dark curve), $L=20$ (dashed curve), $L=30$ (lighter curve). Similarly, the blue curves are built with $\sigma=0.2$ and the red curves with $\sigma=0.9$. In the left panel the null case ($\mathcal{K}=0$) is presented, in the right panel the spacelike case $\mathcal{K}=1$ is presented. In both panels we have set $\kappa=1$ and $x_1=-1$.} 
\label{causc11}
\end{figure}

In order to study the distance between points at the boundary we use equations \eqref{dt} and \eqref{dp}. All results are found numerically, including the turning point. In Fig. \eqref{causc11} we have plotted the quotient $|\Delta t|/|\Delta\phi|$. Every color represents a value $\sigma=0.1,0.5,0.9$. Note that for high values of electrical charge $q>1$ all results lead to a timelike distance between points in the boundary. In both panels ($\mathcal{K}=0$) and  ($\mathcal{K}=1$) the behavior is very similar. As before, note that for $\kappa=+1$ we have that $q=0$.

\section{Light rings in the bulk}\label{sec:5}

We have studied geodesics anchored in the boundary. As we have pointed out, for asymptotically AdS spacetimes there are not timelike geodesics of $Type\,\mathcal{C}$. On the other side, null and spacelike geodesics of $Type\,\mathcal{C}$ can exist, and we have conducted a detailed study for the quBTZ black hole. On this section we are going to link the results on geodesics anchored in the boundary with the existence of a photon sphere. We are going to use a purely geometric method, namely  Gaussian and geodesic curvatures. In order to implement this we build a Riemannian metric by projecting over surfaces of constant energy, the so called Jacobi metric. Recently, new geometric methods have been developed for attacking these kind of problems \cite{Qiao:2022jlu, Qiao:2022hfv, Bermudez-Cardenas:2024bfi,Bermudez-Cardenas:2025duw,Arganaraz:2021fwu}. Let us start with a static metric of the form:

\begin{equation}\label{genm:1}
ds^2=-f(r)dt^2+\frac{dr^2}{h(r)}+h(r)(d\theta^2 +\sin^2(\theta)d\phi^2). 
\end{equation}

A Riemannian metric can be obtained by projecting over surfaces of constant energy $\mathcal{E}$ and momentum $L$, and it is given by \cite{Bermudez-Cardenas:2024bfi}

\begin{equation}\label{jm:2}
J_{ij}dx^{i}dx^{j}=F(r)\left(\frac{dr^2}{f(r)}+h(r)d\theta^2\right)
\end{equation}
where
\begin{equation}\label{Fgen:1}
F(r)=\frac{1}{f(r)}\left(\mathcal{E}^2+\left(\delta-\frac{L^2}{r^2}\right)f(r)\right)
\end{equation}
where $\delta=0,\pm 1$ depending if the trajectory is null, time-like or space-like. The expression in \eqref{jm:2} represents a two dimensional Riemannian metric. In \cite{Qiao:2022jlu} a geometric method for studying the stability of null trajectories $(m=0)$ was developed. In this method, the sign of the Gaussian curvature of the optical metric\footnote{The metric obtained by setting $\delta=0$ in \eqref{jm:2}} together with the vanishing of the geodesic curvature at circular orbits  were used for the classification. Later on, the method was applied to more general metrics including the study of circular null orbits 
\cite{Qiao:2022hfv, Bermudez-Cardenas:2024bfi,Bermudez-Cardenas:2025duw}. The case of timelike orbits was presented in \cite{Bermudez-Cardenas:2024bfi}, where a general expression for the Gaussian curvature of the metrics of the type \eqref{genm:1} was provided. This general expression can be written in terms of the function $f(r)$ of the metric \eqref{genm:1} as:\\
\begin{eqnarray}\label{gauss:1}
\mathcal{K}=\frac{r^2\mathcal{M}(r)}{4E^2 \left(r^2+f \left(\frac{\delta}{E^2} 
   r^2-\sigma^2\right)\right)^3}
   \end{eqnarray}
   where
\begin{eqnarray}
\mathcal{M}(r)&=&-r^4 (f'^2-2ff'')+\sigma^2 f(8f^2+3f'^2 r^2-2rf(rf''+4f'))\nonumber\\
& &\frac{\delta}{E^2} f\left(\left(2f^2f'+8f^3\right)\sigma^2-\frac{\delta}{E^2} 2 f^2 f' r^2-r^3\left(3r f'^2+2ff'+2rff''\right)\right)
\end{eqnarray}
In the massless case, expression \eqref{gauss:1} can be written
\begin{equation}\label{gaussnull:1}
\mathcal{K}=\frac{r^2\left(-r^4 (f'^2-2ff'')+\sigma^2 f(8f^2+3f'^2 r^2-2rf(rf''+4f'))\right)}{4E^2 \left(r^2+f \left(\frac{\delta}{E^2} r^2-\sigma^2\right)\right)^3}
\end{equation}

where we have set $h(r)=r^2$ and $\theta=\pi/2$. Similarly, the geodesic curvature  of an arbitrary circular trajectory defined in \eqref{genm:1} can be written:

\begin{equation}\label{geo:1}
\kappa_g=\frac{2 f(r) -r f'(r)+\frac{2\delta}{E^2}f^2(r)}{E(1+\frac{\delta}{2r E^2}f-\frac{\sigma^2}{r^2})^{3/2}} .
\end{equation}

Thus, a condition for the existence of circular geodesics  can be obtained by the equation $\kappa_g=0$, leading to \cite{Bermudez-Cardenas:2024bfi}
\begin{equation}\label{c:1}
2 f(r) -r f'(r)+\frac{2\delta}{E^2}f^2(r)=0. 
\end{equation}
Condition \eqref{c:1} will provide a criteria for the existence of massive particle surfaces ($\delta\neq 0$) and photon surfaces ($\delta=0$). We are going to use this condition for determining the existence of geodesics connecting a pair of time-like points on the asymptotically AdS boundary.\\
Let us start by showing that the geodesic curvature for null trajectories  can be linked to the existence of geodesics anchored in the boundary. Then, the geodesic curvature $\kappa_g$ transforms to
\begin{equation}\label{geo:null}
\kappa_g=\frac{2 f(r) -r f'(r)}{2rE(1-\frac{\sigma^2}{r^2})^{3/2}},
\end{equation}

\begin{figure}
    \centering
    \includegraphics[width=0.8\linewidth]{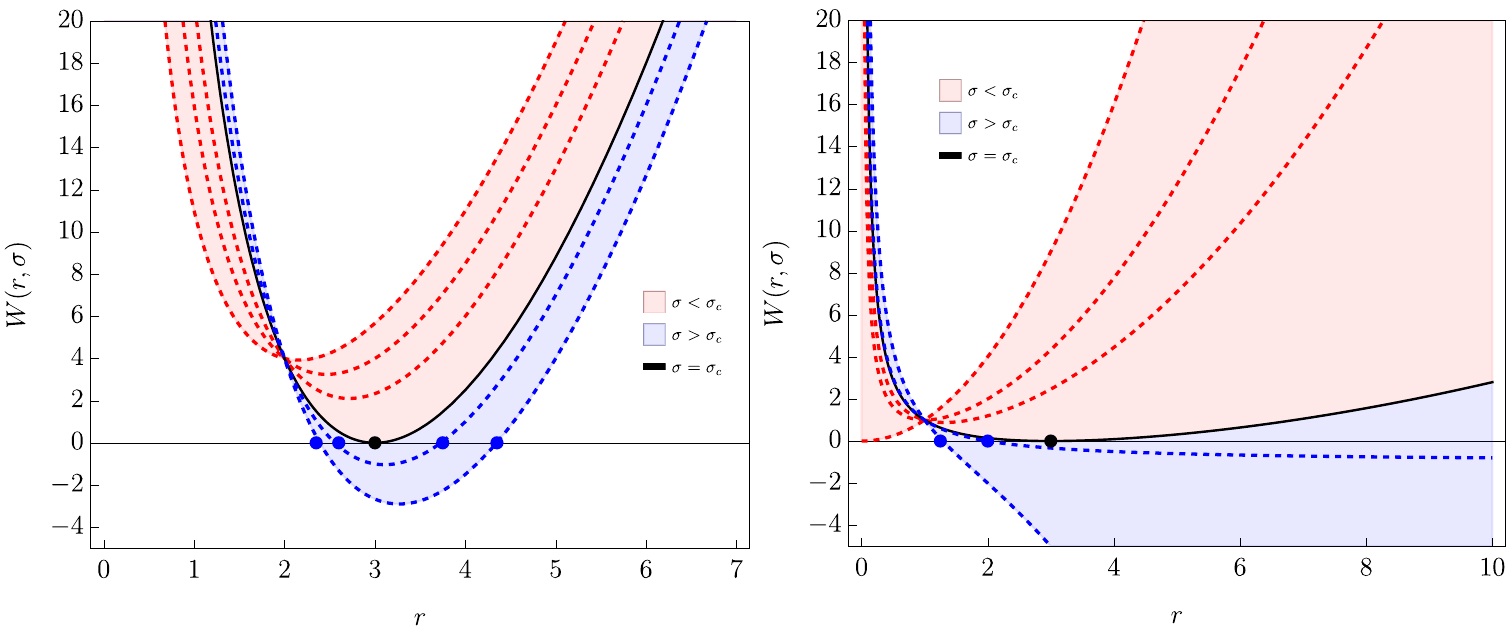}
    \caption{We have plotted the function $W(r,\sigma)=r^2-\sigma^2 f(r) $ for Schwarzschild  (left panel) and Schwarzschild-AdS (right panel) metrics. In both cases we have set $M=1$ The red dashed curves are defined by $\sigma>\sigma_o$, which in the left panel is $\sigma_o=27$ and in the right panel is $\sigma_o=27/28$. The black point in both curvatures represents the only point where a photon sphere is located. The blue dashed curves are defined by $\sigma<\sigma_o$}
    \label{SchW:1}
\end{figure}

Due to the fact that the potential can be written as $V_{\rm eff}=-F(r)f(r)$, with $F(r)$ given by \eqref{Fgen:1}, we have that the condition for a geodesic to have one end at the boundary translates to $\lim_{r\rightarrow \infty}F(r)f(r)>0$. Using \eqref{Fgen:1} the condition for the existence of null geodesics anchored in the boundary is given by

\begin{equation}\label{condC:1}
\lim_{r\rightarrow \infty}F(r)f(r)\approx \lim _{r\rightarrow\infty}\left(1-\frac{\sigma^2}{r^2}f\right)>0.
\end{equation}
The previous condition translates to
\begin{equation}
 \lim_{r\rightarrow \infty}\frac{f}{r^2}<\frac{1}{\sigma^2},
\end{equation}
Once the criteria for the existence of geodesics anchored in the boundary has been defined we have to look for turning points.\\
 In order to have a geodesic of $Type\,\mathcal{C}$, the only extra condition needed is that the effective potential has at least one root, if it has more roots then we chose the biggest one\footnote{The existence of a turning point $r_{t}$ inside the bulk is given by the condition $V_{eff}(r_t)=0$.}. That root $r_t$ could be also  an extrema of the potential, namely $V'(r_t)=0$. When this happens the $r_t$ corresponds to the ubication of a light ring. On the other side, from a pure geometric point of view, the condition for the existence of a light ring is $\kappa_g=0$, from which  we got the well known condition $2 f(r_t) -r_t f'(r_t)=0$. Therefore, if there is a light ring in $r_t$ then the condition for the existence of a turning point is guaranteed\footnote{The argument goes only in one way, if there is not a photon sphere then it does not implies that there is not a turning point.}
The relationship with the photon sphere with the existence of geodesics of $Type\,\mathcal{C}$ was conjectured in \cite{He:2024emd}. The existence of geodesics anchored in the boundary together with the presence of a photon sphere in the bulk leads to the existence of null geodesics of $Type \,\mathcal{C}$. Similarly, in the timelike case the location of a time-like circular orbit is given by the condition \eqref{c:1}, then the existence of a turning point is guaranteed when equation \eqref{c:1} has a solution. Therefore, the existence of the photon sphere constitutes a direct proof of the existence of a turning point. In Fig. \ref{SchW:1} we have plotted the function $W(r,\sigma)=r^2-\sigma^2 f(r)$ for Schwarzschild (left panel)and Schwarzschild-AdS (right panel) black holes.The black point shows the location of the photon sphere, which is over the curve described by $\sigma=\sigma_o$. Note how the location of the roots, and therefore of the turning point, changes  when we consider $\sigma<\sigma_o$ (red curves) and  $\sigma>\sigma_o$ (blue curves). In the Schwarzschild case it is clear that there is only one curve with a minima located at $W(r,\sigma)=0$, any other curve has two roots or none. In the Schwarzschild-AdS case, a few curves have only one root. Nevertheless, the only curve that has a root which is a minimum of the function $W(r,\sigma)$ is given by $\sigma=\sigma_o$, and is located over the photon sphere.\\
When we set the geodesic curvature to zero we obtain a family of photon sphere radii parametrized by $\sigma$, and we would like to know what happens if we move outside the photon sphere determined by a a fixed $\sigma_o$. The location of a photon sphere with parameter $\sigma_o$ is given by $r_{pho}=\sigma_o^2f(r_{pho})$ and as we move infinitesimally from $r_{pho}$  the turning point transforms to

\begin{equation}\label{turnp:1}
    r_{t}(\sigma)=r_{pho}+\sqrt{\frac{4r_{pho}f(r_{pho})(\sigma f^{1/2}(r_{pho})-r)}{1-r_{pho}^2f''(r_{pho})}}+\mathcal{O}(\sigma-\sigma_o)
\end{equation}

The last equation shows that for any value $\sigma>\sigma_o$ the resulting returning point is located outside the photon sphere, in other words, the turning point now is bigger than the photon sphere radius. See appendix \ref{App:3} for a detailed calculation. 

\subsection{Asymptotically AdS spacetimes without light rings}
In pure AdS spacetime, the blackening function is \eqref{ads} and the condition \eqref{condC:1} is satisfied, when $\ell^2>\sigma^2$. Thus, the existence of null geodesics of $Type\,\mathcal{C}$  is guaranteed provided that there are geodesics anchored in the boundary.

The sign of $\mathcal{K}$ will determine the existence of geodesics with at least one end at the boundary. Thus, when $\mathcal{K}<0$ (time-like geodesics) there cannot be timelike geodesics anchored in the boundary, thereby the only possibilities remaining are null and space-like geodesics.

When $\mathcal{K}=0$ there are going to be geodesics anchored in the boundary. Now let see what happens with the geodesic curvature:
\begin{equation}
\kappa_{AdS}=\frac{\mathcal{E}^2r}{(\mathcal{E}^2-m^2(1+r^2))^{3/2}}.
\end{equation}
The only possibility is that the light ring is located at infinity, therefore there is not a photon sphere that enforces a geodesic anchored in the boundary to return to the boundary. However, the point $r_{t}$ is the largest root of $V_{eff}(r)$ and is given by  $r_t=\sigma/\sqrt{1-\sigma^2}$, which implies that there will be null geodesics of the $Type\,\,\mathcal{C}$. This is a clear example that the absence of a photon sphere does not mean that there is not a turning point.

In three dimensions, an asymptotically AdS black hole  is the BTZ black hole
\begin{equation}
ds^2=-\left(-M+\frac{r^2}{\ell_{3}^2}\right)dt^2+\frac{dr^2}{\left(-M+\frac{r^2}{\ell_{3}^2}\right)}+r^2 d\phi^2.
\end{equation}

The geodesic curvature of  a circular orbit of the Jacobi for the BTZ black hole can be written
\begin{equation}
\kappa_{gBTZ}=-\frac{M+\frac{m^2}{\mathcal{E}^2}\left(-M+\frac{r^2}{\ell_3^2}\right)^2}{\mathcal{E}\left(1-\frac{m^2}{\mathcal{E}^2}\left(-M+\frac{r^2}{\ell^2_3}\right)\right)^{3/2}}.
\end{equation}

The previous geodesic curvature vanishes when calculated for circular orbits leading to a master equation that helps us to determine if a massive particle surface is present, indeed
\begin{equation}\label{masterbtz}
\frac{E^2}{m^2}=-\frac{M}{\left(-M+\frac{r^2}{\ell^2_3}\right)^2}.
\end{equation}

The previous condition shows that there are not massive particle surfaces around the BTZ black hole without rotation. Similarly, when $m=0$ the geodesic curvature transforms to $\kappa_g=-\frac{M}{\mathcal{E}}$ which shows that light rings do not exist. On the other side, Gaussian curvature \eqref{gauss:1} becomes:

\begin{equation}\label{gaussian:1}
\mathcal{K}_{BTZ}=\frac{\mathcal{E}^4\ell^2_3 M+m^2\left(m^2\ell_{3}^2\left(\frac{r^2}{\ell_{3}^2}-M\right)^3-3\mathcal{E}r^2\left(\frac{r^2}{\ell_3^2}-M\right)\right)}{\ell_3^{4}\left(\mathcal{E}^2-m^2\left(\frac{r^2}{\ell_3^2}-M\right)\right)^3}.
\end{equation}

The Gaussian curvature \eqref{gaussian:1} is negative at infinity, thus $\lim_{r\rightarrow \infty}\mathcal{K}_{BTZ}=-1/(m^2\ell_{3}^2)$,
which shows that at the boundary the geometry of the Jacobi metric has constant negative curvature. At infinity, the bigger the mass the smaller the curvature, therefore the Gaussian curvature goes to $-\infty$  from negative values. When  $r\rightarrow 0$ the Gaussian curvature becomes
\begin{equation}
\lim_{r\rightarrow 0}\mathcal{K}_{BTZ}=-\frac{M\ell^2_3(\mathcal{E}^2+m^2M^2)}{\ell^4(\mathcal{E}+m^2M)^3}.
\end{equation}
Both limits show that the Gaussian curvature is negative and there is not a change in between, a similar behavior is presented by the geodesic curvature. It is clear that when $r\rightarrow\infty$ the Gaussian curvature $\left.\mathcal{K}_{BTZ}\right|_{m=0} $ remains constant and negative. Near the horizon we have 
$\lim_{r\rightarrow r_{H}}\mathcal{K}_{BTZ}=-M/(\mathcal{E}^2\ell^2_3)$. This shows that for null trajectories the Gaussian curvature never vanishes, it is constant and negative everywhere, and therefore any orbit, if it exists, it remains unstable.\\

\subsection{The quBTZ black hole}
Replacing \eqref{qBTZ:1} in the expression of geodesic curvature \eqref{geo:1} and equating it to zero we obtain the following restriction
\begin{equation}\label{qubtzcond}
16 \mathcal{G}_3 M+\frac{3\ell \mathcal{F}(M)}{r}+\frac{2m^2\left(8 \mathcal{G}_3 M+\frac{\ell \mathcal{F}(M)}{r}-\frac{r^2}{\ell_3^2}\right)^2}{\mathcal{E}^2}=0
\end{equation}
When $\ell=0$ we recover the BTZ case condition \eqref{masterbtz}:
\begin{equation}
16 \mathcal{G}_3 M+\frac{2m^2\left(8 \mathcal{G}_3 M-\frac{r^2}{\ell_3^2}\right)^2}{\mathcal{E}^2}=0
\end{equation}

The equation \eqref{qubtzcond} has no roots and therefore there are not timelike geodesics. When $m=0$ we get
\begin{equation}
r=-\frac{3\ell F(M)}{16\mathcal{G}_3M}
\end{equation}

The sign of the previous equation shows that there is a light ring when $M<0$. From equation \eqref{mqbtz} we can deduce that in order to have a negative value of $M$ we must have $\kappa x_1^2<1$, which implies that $\kappa=+1$ and $0<x_1<1$. Therefore, the so called branch 1a of the quBTZ allows the existence of light rings. Therefore the exitence of null geodesics anchored in the bundary is guaranteed. In this branch the lower bound is reached for $\kappa=+1$ and $x_1=1$, and the renormalized stress tensor vanishes. Note that the metrics with negative masses correspond, in the limit $\ell\rightarrow 0$, to conical singularities in $AdS_3$.

\section{Discussion and final comments}\label{sec:6}
We have analyzed bulk probes of the quBTZ black holes including the charged case. We have found the conditions needed for the existence of geodesics of $Type,\mathcal{C}$. We have made an extensive anlaysis for different cases and we have completed examples in all branches of  the quBTZ.  As it happens in asymptotically AdS spacetimes there is not time-like geodesics whith endpoints in the boundary, therefore we focused our analysis in null and space-like geodesics. Several cases where geodesics of $Type \ \mathcal{C}$ can be found where detected and therefore a calculation of entanglement entropy can be carried out. An important aspect is the type of separation between two points in the boundary. The distance between points in the boundary, which in spite of being in a two dimensional manifold are difficult to calculate analytically,  have been performed using numerical methods.\\
We have also gave some light to the conjecture claiming that the existence of a photon sphere will lead to the existence of a turning point, and therefore a geodesic anchored in the boundary will come back, maybe after one or more spins, to the boundary \cite{He:2024emd}. The result was confirmed for a general static spherically symmetric metric. The result is a corollary of another interesting fact, the geodesic curvature of curves defined over a surface of constant energy can be used to determine the existence of light rings or timelike circular orbits. If a black hole has a light ring, then the geodesic anchored in the boundary will turn back to the boundary. \\
It is known that extremal surfaces cannot be continuously deformed past any surface with nonpositive extrinsic curvature. Moreover, the outermost surface  has nonnegative extrinsic curvature \cite{Engelhardt:2013tra}. Using intrinsic properties such a Gaussian and geodesic curvatures we can also attack the barrier problem described before using the results comming from Riemannian geometry. In that sense, extremal probes can be studied using geometric tools presented in this article. We let this for future work. 

\appendix
\section{Parameter-space conditions for turning points of $V_{\mathrm{eff},5}$}\label{App:1}

We now present an analytic characterization of the turning points, which allows us to determine the region in the $(\sigma,L)$ parameter space where a real, positive turning point exist. The turning points
$r_t$ are determined by the real positive roots of the quintic equation \eqref{pol:2},
which we write explicitly as
\begin{equation}\label{pol:2_app}
r^5-r^3 \left(1-\frac{1}{\sigma^2}\right)L^2-\frac{8}{27}r^2+\frac{8}{27}L^2=0.
\end{equation}

Although quintic equations of this type can in principle be reduced analytically, e.g. through standard Tschirnhaus reductions to principal or Bring--Jerrard form, such closed-form representations are not especially useful for selecting the physical turning radius \cite{AdamchikJeffrey:2003}. It is therefore more convenient to determine directly the ranges of $L$ for which a real positive turning point $r_t$ exists. Thus, we rewrite \eqref{pol:2_app} as
\begin{equation}\label{Pr_def}
P(r)=r^5-\Delta r^3-\frac{8}{27}r^2+\frac{8}{27}L^2.
\end{equation}
Differentiating equation \eqref{Pr_def}, we obtain
\begin{equation}\label{derivP}
P'(r)=r\left(5r^3-3\Delta\, r-\frac{16}{27}\right),
\end{equation}
with
\begin{equation}\label{Delta_def}
\Delta=L^2\left(1-\frac{1}{\sigma^2}\right).
\end{equation}

For $r>0$ we have $P'(r)=r\, Q(r)$ with
\begin{equation}
    Q(r)=5r^{3}-3\Delta r-\frac{16}{27}.
\end{equation}
Since $Q(0)<0$ and $Q(r)\to+\infty$ as $r\to\infty$, the equation $Q(r)=0$
admits at least one positive solution. Moreover, since the coefficients of $Q(r)$ exhibit a single sign change for any $\Delta \neq 0$, the vanishing coefficient can be discarded, and the remaining sequence still shows one sign change. Hence, by Descartes's rule, $Q(r)$ has exactly one positive root. Hence there exists a unique $r_\star>0$ such that
\begin{equation}\label{Qr}
5r_{*}^{3}-3\Delta r_{*}-\frac{16}{27}=0.
\end{equation}
Therefore $P'(r)=rQ(r)$ changes sign only once on $(0,\infty)$, so $P(r)$ has a single critical point $r_\star$ in $(0,\infty)$, which is the global minimum of $P(r)$ for $r>0$. Turning points correspond to real positive roots of $P(r)=0$, equivalently of $V_{\mathrm{eff}}(r)=0$ for $r>0$, since $P(r)=-L^{2}r^{3}V_{\mathrm{eff}}(r)$, and the outer
turning radius is the largest such root. Moreover, $P(0)>0$ and $P(r)\to+\infty$ as $r\to\infty$. Since $r_\star$ is the unique minimum,
there are three possibilities: if $P(r_\star)>0$ then $P(r)>0$ for all $r>0$ and there is no
turning point; if $P(r_\star)=0$ then $P$ has a double positive root at $r_\star$ (critical case);
and if $P(r_\star)<0$ then $P$ has exactly two positive roots, corresponding to an inner and an outer turning radius.\\
Evaluating $P(r)$ at the horizon one finds
\begin{equation}\label{P_at_rh}
P(r_h)=P\!\left(\frac{2}{3}\right)=\frac{8}{27}\left(L^2-\Delta\right)=\frac{8}{27}\frac{L^2}{\sigma^2}>0,
\end{equation}
so a turning point outside the horizon requires not only $P(r_\star)<0$ (two positive roots), but also that the minimum occurs at $r_\star>r_h$; otherwise both roots lie at $r<r_h$.

This latter condition can be written explicitly evaluating \eqref{Qr} at $r_h=\frac{2}{3}$:
\begin{equation}
5r_h^3-3\Delta r_h-\frac{16}{27}
=\frac{8}{9}-2\Delta.
\end{equation}
Since the left-hand side of \eqref{Qr} is negative at $r=0$ and grows to $+\infty$ for large $r$, we get
\begin{equation}\label{rstar_vs_rh}
r_\star>r_h \quad \Longleftrightarrow \quad \Delta>\frac{4}{9}.
\end{equation}
Let us solve the system $P(r)=0$ and $P'(r)=0$. At parameter values where the number of
positive roots of $P(r)$ changes, $P(r)$ admits a positive root of multiplicity two at
$r=r_\star>0$, i.e. $P(r_\star)=0$ and $P'(r_\star)=0$. Equivalently, $P$ and $P'$ share a common root, which holds
if and only if the resultant vanishes, $Res(P,P')=0$ (see e.g.
\cite{GKZ:Resultants,Ahmedov:2021universe}). Computing the
resultant explicitly (via the Sylvester matrix), we obtain
\begin{equation}
Res(P,P')=-(108)^4 L^2 \mathcal{P}(\Delta,L),
\end{equation}
where 
\begin{align}
\mathcal{P}(\Delta,L)&=531441 \Delta^5 L^2
-6912 \Delta^3-356400\Delta ^2 L^2+ \nonumber\\
&\quad+1620000 \Delta L^4-
1350000 L^6+4096.
\end{align}

Note that $P$ has a unique critical point $r_\star>0$. Therefore, the condition
$\mathcal{P}(\Delta,L)=0$ marks the threshold where the global minimum satisfies $P(r_\star)=0$,
i.e.\ where $P$ acquires a double positive root. Consequently, the sign of $\mathcal{P}(\Delta,L)$
distinguishes the two-root regime ($P(r_\star)<0$) from the no-root regime ($P(r_\star)>0$). Along the physical slice
$\Delta=L^2(1-\sigma^{-2})$, $\mathcal{P}$ therefore determines the allowed ranges of $(\sigma,L)$ for which
turning points exist; imposing \eqref{rstar_vs_rh} further selects those with $r_t>r_h$. \\
Let us analyze three different cases for the values of $\sigma$.

\paragraph{Case $\sigma<1$.}
If $\sigma<1$, then
$\Delta=L^2\left(1-\frac{1}{\sigma^2}\right)<0$.
In this regime, the condition for the existence of a double root is
\begin{equation}
\mathcal{P}\!\left( 
\Delta,L_c\right)=0.
\end{equation}
where $L_c$ is a critical value of $L$ at which $P(r)$ acquires a positive root of multiplicity two. Substituting $\Delta=L^2\left(1-\frac{1}{\sigma^2}\right)$ into $\mathcal{P}(\Delta,L)$, one obtains a polynomial in $x=L^6$:
\begin{equation}\label{Psigma_def}
\mathcal{P}_{\sigma}(L)=a(\sigma)x^2+b(\sigma)x+c,\qquad x=L^6,
\end{equation}
with
\begin{eqnarray}
a(\sigma)&=&531441 \left(1-\frac{1}{\sigma^2}\right)^5,\nonumber\\
b(\sigma)&=&-6912\left(1-\frac{1}{\sigma^2}\right)^3-356400\left(1-\frac{1}{\sigma^2}\right)^2
+1620000\left(1-\frac{1}{\sigma^2}\right)-1350000,\nonumber\\
c&=&4096.\nonumber
\end{eqnarray}
Since $a(\sigma)<0$ and $c>0$, the quadratic in \eqref{Psigma_def} has exactly one positive root, which we denote by $x_*(\sigma)>0$, and we define
\begin{equation}
L_c\equiv L_{\max}(\sigma)=\big(x_*(\sigma)\big)^{1/6}.
\end{equation}
Furthermore, $\mathcal{P}_\sigma(0)=c>0$ and $\mathcal{P}_\sigma(L)\to -\infty$ as $L\to\infty$, hence
\begin{equation}
\mathcal{P}_\sigma(L)>0 \ \Longleftrightarrow\ 0\ <  L<L_{\max}(\sigma),
\qquad
\mathcal{P}_\sigma(L)<0 \ \Longleftrightarrow\ L>L_{\max}(\sigma).
\end{equation}
However, since $\Delta<0<\frac{4}{9}$, Eq.~\eqref{rstar_vs_rh} implies $r_\star<r_h$, and together with \eqref{P_at_rh} this shows that, even when $\mathcal{P}_\sigma(L)>0$ and two positive roots exist, both of them lie inside the horizon. Therefore, there is no turning point with $r_t>r_h$ for $\sigma<1$.

\paragraph{Case \texorpdfstring{$\sigma=1$}{sigma=1}.}
If $\sigma=1$ we have $\Delta=0$, and thus
\begin{equation}
\mathcal{P}(0,L)=4096-1350000\,L^6.
\end{equation}
The critical value $L=L_c$ is determined by $\mathcal{P}(0,L_c)=0$, namely
\begin{equation}
\mathcal{P}(0,L)>0 \ \text{iff}\ 0 < L<L_c,
\qquad
L_c=\left(\frac{256}{84375}\right)^{1/6}\approx 0.3805.
\end{equation}
At $L=L_c$ the polynomial $P$ develops a double positive root at its minimum $r_\star$, while for
$0< L<L_c$ continuity implies $P(r_\star)<0$. Since $P$ has a unique minimum for $r>0$, it follows that
$P$ has exactly two positive roots. For $\Delta=0$, using \eqref{Qr} the minimum is located at
\begin{equation}
r_\star=\left(\frac{16}{135}\right)^{1/3}<\frac{2}{3}.
\end{equation}
Moreover, by \eqref{P_at_rh} one has $P(r_h)>0$. Because $P$ is strictly increasing for $r>r_\star$ and $r_h>r_\star$,
the larger positive root must satisfy $r<r_h$ (while the smaller one is already $<r_\star$). Therefore both positive
roots lie below $r_h$, and there is no turning point with $r_t > r_h$ in the case $\sigma=1$.

\paragraph{Case $\sigma>1$.}
If $\sigma>1$, then 
$\Delta=L^2\left(1-\frac{1}{\sigma^2}\right)>0$.
In this regime $\mathcal{P}\!\left(\Delta, L\right)$ can again be written as a quadratic polynomial in $x=L^6$ of the form \eqref{Psigma_def} with the same coefficients $a(\sigma),b(\sigma),c$.
When the discriminant $b(\sigma)^2-4a(\sigma)c$ is positive, we denote its two real roots by $x_\pm(\sigma)$ and define
\begin{equation}
L_\pm(\sigma)=\big(x_\pm(\sigma)\big)^{1/6}.
\end{equation}
Since $a(\sigma)>0$ for $\sigma>1$, we have $\mathcal{P}_\sigma(L)>0$ for $L^6<x_-(\sigma)$ or $L^6>x_+(\sigma)$, and $\mathcal{P}_\sigma(L)<0$ for $x_-(\sigma)<L^6<x_+(\sigma)$.
Hence, two positive roots of $P(r)$ exist precisely when $\mathcal{P}_\sigma(L)>0$.

Finally, in order for the turning point to be outside the horizon we must also impose \eqref{rstar_vs_rh}, namely
\begin{equation}
\Delta=L^2\left(1-\frac{1}{\sigma^2}\right)>\frac{4}{9}
\quad \Longleftrightarrow \quad
L>\frac{2}{3}\left(1-\frac{1}{\sigma^2}\right)^{-1/2}.
\end{equation}
Therefore, for $\sigma>1$ the condition for the existence of a turning point with $r_t>r_h$ is the simultaneous validity of
\begin{equation}
\mathcal{P}_\sigma(L)>0,
\qquad
L>\frac{2}{3}\left(1-\frac{1}{\sigma^2}\right)^{-1/2}.
\end{equation}

\section{The turning point and the photon sphere} \label{App:3}
As we have discussed, the existence of a photon sphere ensures the existence of a turning point.  That point is going to be located over the photon sphere. However, we would like to know what happens with the turning point if we move out from the photon sphere. We start by defining a two-variable function in the following way:
\begin{equation}\label{wfunc:1}
W(r,\sigma)=r^2-\sigma^2 f(r).
\end{equation}
The function $W(r,\sigma)$ has a zero at $r_{ph}$ the location of the photon sphere. That root is unique because it represents a global minimum of $W(r,\sigma)$. Thus, the following system of equations is satisfied at the photon radius $r_{ph}$:
\begin{eqnarray}
  W(r_{ph},\sigma)&=&0\label{wcond:1}\\
  \partial_{r}W(r,\sigma)\rvert_{r_{ph}}&=&0\label{wcond:2}
\end{eqnarray}
The value of $r_{ph}$ depends on $\sigma$ therefore on momenta $L$ and energy $E$. Let us fix the value of $\sigma$ to $\sigma_o$ then the radius of the photon sphere is fixed to the value 
\begin{equation}\label{rpho:1}
r_{pho}=\sigma^2_of(r_{pho}).  
\end{equation}
Replacing \eqref{rpho:1}  in \eqref{wfunc:1} we obtain
\begin{equation}
    W(r_{pho},\sigma)=f(r_{pho})(\sigma_{o}^2-\sigma^2).
\end{equation}
Therefore, if $\sigma^2>\sigma_{o}^2$ then $W(r_{pho},\sigma)<0$, which in its turn leads to the existence of at least one more root. In other words, although  $\partial_r W(r,\sigma)=0$, for $r=r_{pho}$ is satisfied, the condition $W(r_{pho},\sigma)=0$  is not satisfied anymore.\\
We want to analyze what happens with the returning point when we move infinitesimally from $\sigma_o$. We will find an expression for the turning point for small deviations of $\sigma_o$. In order to do that we are going to perform a series expansion around the critical point. We start by introducing the notation $\Delta r=r-r_{pho},\,\,\Delta\sigma=\sigma-\sigma_o$, then a series expansion around $(r_o,\sigma_o)$ is given by
\begin{equation}
    \begin{split}        
W(r,\sigma)&=W(r_{pho},\sigma_o)+\partial_r W(r_{pho},\sigma_o)
\Delta r+\partial_\sigma W(r_{pho},\sigma_o)\Delta\sigma\\
& 
+\frac{1}{2}\partial_{rr}W(r_{pho},\sigma_o)(\Delta r)^2+\frac{1}{2}\partial_{\sigma\sigma}W(r_{pho},\sigma_o)(\Delta \sigma)^2+\partial_{r\sigma}W(r_{pho},\sigma_o)\Delta r\Delta \sigma\\
& +\mathcal{O}(\lvert(\Delta r, \Delta \sigma)\rvert^3 ). 
\end{split}
\end{equation}
Using \eqref{wcond:1} and \eqref{wcond:2} the previous expression transforms to 
\begin{equation}\label{genW:1}
\begin{split}
   W(r,\sigma)=1-\frac{r_{pho}^2}{2f(r_{pho})}f''(r_{pho})(\Delta r)^2-2\sigma_o f(r_{pho}) \Delta(\sigma)-f(r_{pho})(\Delta \sigma)^2-2\sigma_o f'(r_{pho})\\
   -2\sigma_o f'(r_{pho})\Delta r\Delta \sigma+\mathcal{O}(\lvert(\Delta r, \Delta \sigma)\rvert^3 ).
   \end{split}
\end{equation}
On the other hand, according to the implicit function theorem, there exist a unique function $h(r)$ such that
\begin{equation}
    \sigma=h(r),\,h(r_{pho})=\sigma_o,\, W(r,h(r))=0,\,\,\forall r\sim r_{pho}
\end{equation}
therefore
\begin{equation}
    h'(r_{pho})=\frac{\partial_r W}{\partial _\sigma W}(r_{pho},\sigma_o)=0,\, h''(r_{pho})=-\frac{\partial_{rr}W}{\partial_\sigma W}(r_o,\sigma_{pho})=\frac{2-\sigma_o^2 f''(r_{pho})}{2\sigma_o f(r_{pho})}.
\end{equation}
Finally, we can write
\begin{eqnarray}
    \Delta \sigma&=&\frac{1}{2}h''(r_{ph})(\Delta r)^2+\mathcal{O}((\Delta r)^3)\, ,
\end{eqnarray}
hence, from the previous equation we find
\begin{equation}
\Delta r=\mathcal{O}(\sqrt{\Delta \sigma}).
\end{equation}
Thus, a general solution for $\Delta r$ has to be of the form
\begin{equation}\label{ans:1}
    \Delta r=x(\Delta \sigma)^{1/2}+\mathcal{O}(\Delta \sigma), \, \Delta \sigma>\sigma_o
\end{equation}
Replacing \eqref{ans:1} in \eqref{genW:1} and using \eqref{rpho:1}  we obtain
\begin{equation}
    x=\pm\sqrt{\frac{2 r_{pho}f^{1/2}(r_{pho})}{1-\frac{r_{pho}^2}{2f(r_{pho})}f''(r_{pho})}}.
\end{equation}
Finally, from expression \eqref{ans:1} we get \eqref{turnp:1}.

\section{The general case}\label{App:2}

For null geodesics $(\kappa=0)$ the previous potential becomes
\begin{equation}
V_{eff}=\left(\frac{1}{\ell^2_3}-\frac{8\mathcal{G}_3 M}{r^2}-\frac{\ell F(M)}{r^3}\right).
\end{equation}
Therefore when $r\rightarrow \infty$ we have $\lim_{r\rightarrow\infty}V_{eff}(r)=1/\ell^2_3$, 
hence the potential $V_{eff}$ in \eqref{geor} becomes
\begin{equation}
\dot{r}^2_{\infty}=\frac{1}{\sigma^2}-\frac{1}{\ell_3^2}\geq 0, 
\end{equation}

therefore there are null geodesics with at least an end point if $ |\sigma|\leq \ell_3$. \\
When $\kappa=1$  at the limit $r\rightarrow \infty$, we get $V(r)\rightarrow 1/\ell_3^2$, then there is at least one endpoint at the boundary. Moreover, when $\dot{r}=0$ a turning point should exists, therefore
\begin{equation}\label{nullr:1}
W(r)=r^5\left(\frac{1}{\sigma^2}-\frac{1}{\ell_3^2}\right)+8 \mathcal{G}_3M+\ell F(M)=0.
\end{equation}

We want to know if the previous equation has solution. There are two particular cases \footnote{The trivial case $\sigma^2=\ell_{3}^2$ leads to the condition $8\mathcal{G}_3M>0$ which is satisfied for $M>0$.}, the first cases is given by $|\sigma|< |\ell_3|$, then $W'(r)>0$ which shows that the function is monotonically increasing  , and because $\lim_{r\infty}W(r)=+\infty$ and $\lim_{r\rightarrow 0}=\ell F>0$ we conclude that the function $W(r)$ never reaches negative values, then the equation \eqref{nullr:1} does not have a solution which implies that the geodesic does not return to the boundary. The second case $|\sigma|>\ell_3$ leads to a condition for the existence of a solution of \eqref{nullr:1}. Because $\lim_{r\rightarrow \infty}=-\infty$ and $\lim_{r\rightarrow 0}=\infty$ there should be a point where the function $W(r)$ reaches zero, this will happen only when
\begin{equation}
r>\frac{\sigma^{1/2}\ell_{3}^{1/2}}{5^{1/4}(\sigma^2-\ell_3^2)^{1/4}}.
\end{equation}

Once we have determined that there exist a geodesic with two end points at the boundary we want to see if the separation between the points is light-like.\\
From equations \eqref{dt} and \eqref{dp} we find
\begin{eqnarray}
\frac{dt}{dr}&=&\frac{r}{\left(\frac{r^2}{\ell_3^2}-\frac{8\mathcal{G}_3M}{r}\right)\sqrt{r^2-\sigma^2\left(\frac{r^2}{\ell_3^2}-\frac{8\mathcal{G}_3M}{r}\right)}},\\
\frac{d\phi}{dr}&=&\frac{\sigma}{r\sqrt{r^2-\sigma^2\left(\frac{r^2}{\ell_3^2}-\frac{8\mathcal{G}_3M}{r}\right)}}.\\
\end{eqnarray}
The solutions to previous equations are given by
\begin{eqnarray}
t(r)&=&-\frac{2r^3\sqrt{\left(1+\frac{(\ell_3^2-\sigma^2)r^3}{\ell_3^2\sigma^2 8 \mathcal{G}_3M}AF_1\left(\frac{7}{6},\frac{1}{2},1,\frac{13}{6},\frac{(\sigma^2-\ell_3^2)r^3}{8\ell_3^2 \sigma^2\mathcal{G}_3 M},\frac{r^3}{\ell F(M)\ell_3^3}\right)\right)}}{7\ell_3 F(M)\sqrt{r^2+\frac{\sigma^2\left(8\ell^2_3\mathcal{G}_3M-r^3\right)}{\ell_3^2r}}}\\
\phi(r)&=&\frac{2r\sqrt{r^2}+\frac{\sigma^28\mathcal{G}_3M\ell^2_3-r^3}{\ell_3^2r}H_2F_1\left(\frac{2}{3},1,\frac{7}{6},\frac{(\sigma^2-\ell_3^2)r^3}{8\ell_3^2 \sigma^2\mathcal{G}_3 M}\right)}{8\sigma\mathcal{G}_3 M }
\end{eqnarray}

where $AF_1(a,b^1,b^2,c,x,y)$ is the Appell hypergeometric function of two variables $x,y$, and $H_2F_1$ is the hypergeometric function. The previous expression are cumbersome and depend on several parameters.  Because we are interested in the calculating distance between points in the boundary we can take $\ell_3=1$. As we have done in previous sections.

\end{document}